\documentclass[graybox]{svmult}
\usepackage{mathptmx}       
\usepackage{helvet}         
\usepackage{tikz}
\usepackage{ifthen}
\usetikzlibrary{arrows.meta,positioning,shapes,backgrounds,calc}
\usetikzlibrary{arrows.meta,positioning}
\usepackage{courier}        
\usepackage{type1cm}        
                        \usepackage{comment} 
\usepackage{todonotes} 
\usepackage[ruled,vlined]{algorithm2e} 
\usepackage{makeidx}         
\usepackage{graphicx}        
\usepackage{multicol}        
\usepackage[bottom]{footmisc}
\usepackage{amsmath} 

\usepackage{amsmath} 
\newcommand{\BR}{\ensuremath{\mathrm{BR}}}
\usepackage{amsmath}
\usepackage{amssymb}
\usepackage{amscd}
\usepackage{indentfirst}
\usepackage{tabularx}
\usepackage{multirow}

\def\1{{\bf 1}}

\makeatletter
\newcommand\para{\@startsection{paragraph}{4}{\z@}%
  {3.25ex \@plus1ex \@minus.2ex}
  {-1em}
  {\normalfont\normalsize\bfseries}}
\makeatother

\makeindex             

\begin{document}

\title*{Toward a Multi-Echelon Cyber Warfare Theory: A Meta-Game-Theoretic Paradigm for Defense and Dominance}
\titlerunning{Multi-Echelon Cyber Warfare Theory}
\author{Ya-Ting Yang and Quanyan Zhu}
\authorrunning{Y.T. Yang and Q. Zhu}
\institute{Department of Electrical and Computer Engineering, Tandon School of Engineering, New York University\\ \email{yy4348@nyu.edu, qz494@nyu.edu} }

\maketitle

\abstract{Cyber warfare has become a central element of modern conflict, especially within multi-domain operations. As both a distinct and critical domain, cyber warfare requires integrating defensive and offensive technologies into coherent strategies.  While prior research has emphasized isolated tactics or fragmented technologies, a holistic understanding is essential for effective resource deployment and risk mitigation. Game theory offers a unifying framework for this purpose. It not only models attacker-defender interactions but also provides quantitative tools for equilibrium analysis, risk assessment, and strategic reasoning.  Integrated with modern AI techniques, game-theoretic models enable the design and optimization of strategies across multiple levels of cyber warfare, from policy and strategy to operations, tactics, and technical implementations.  These models capture the paradoxical logic of conflict, where more resources do not always translate into greater advantage, and where nonlinear dynamics govern outcomes.  To illustrate the approach, this chapter examines \emph{RedCyber}, a synthetic cyber conflict, demonstrating how game-theoretic methods capture the interdependencies of cyber operations. The chapter concludes with directions for future research on resilience, cross-echelon planning, and the evolving role of AI in cyber warfare.}


\section{Cyber Warfare: Definition and Key Characteristics}

Cyber warfare refers to the strategic deployment of cyber attacks by nation-states or organized entities to disrupt, damage, or seize control of an adversary's digital and physical infrastructure. These actions are designed to undermine national security, disrupt civil infrastructure, and destabilize economies. Common targets include government networks, financial institutions, and essential utilities, all of which are critical for the functioning of modern societies.

The primary targets of cyber warfare are systems and services that play a vital role in both the public and private sectors. These include government networks and military systems, critical infrastructure such as power grids and water supply systems, and essential civilian services such as healthcare and communications. By focusing on these targets, attackers aim to cause widespread disruption and chaos.

Cyber warfare employs a variety of sophisticated methods to achieve its objectives. Malware attacks, including viruses, ransomware, and trojans, are commonly used to infiltrate and compromise systems. Distributed Denial of Service (DDoS) attacks, which overwhelm servers with excessive traffic, are employed to disable systems and services. Cyber espionage, another key tactic, involves the theft of sensitive information such as military intelligence or corporate secrets. In some cases, hybrid tactics are used, combining cyber operations with physical military actions to enhance the overall impact.

The impact of cyber warfare can be far-reaching and devastating. Essential services like energy, transportation, and healthcare can be disrupted, causing significant social and economic damage. Financial systems may be destabilized, resulting in widespread economic consequences. Additionally, societal disinformation campaigns can undermine public trust and cohesion, further exacerbating the damage caused by these attacks. As a result, cyber warfare has emerged as a critical threat in modern conflicts, demanding robust prevention and mitigation strategies.

\subsection{The Current State of Cyber Warfare}

Today’s cyber warfare is characterized by escalating threats from both nation-states and non-state actors, including hacktivists, terrorist organizations, and organized crime groups. These entities increasingly deploy cyber tactics to exert power, disrupt systems, and achieve their objectives. Prominent examples include North Korea's Lazarus Group, which has targeted cryptocurrency exchange platforms, causing substantial financial losses and exposing vulnerabilities in digital financial systems \cite{bbc2025bybit}. Russia's use of DDoS attacks during the Georgian conflict in 2008 is an example of how cyber tools can be abused during geopolitical tensions \cite{connell2016russia}. Additionally, Stuxnet remains a landmark case of a nation-state using cyber warfare to disrupt critical infrastructure, specifically targeting Iran's nuclear enrichment program \cite{farwell2011stuxnet}.

Modern warfare encompasses multiple domains, including air, space, land, sea, and cyberspace, requiring integrated strategies that leverage the unique advantages of each. Cyber warfare has become a pivotal component of these multi-domain military strategies. A notable example is Israel's Operation Orchard in 2007, where cyber capabilities were employed to misinform Syrian air defenses, facilitating a successful airstrike on a nuclear facility \cite{abcnews2007syria}. In such multi-domain operations, it is crucial to strategically plan cyber and non-cyber operations to ensure that each action amplifies the advantages created in other domains. This cross-domain operational planning is essential in modern cyber warfare, enabling forces to maintain a strategic edge over adversaries.

One of the significant characteristics of cyber warfare is anonymity and the difficulty in attributing attacks to specific perpetrators. The anonymous and decentralized nature of cyberspace allows attackers to obfuscate their identities, complicating efforts to identify and respond to threats. This lack of attribution not only delays timely responses but also aggravates geopolitical tensions, as nations struggle to determine whether an attack originates from a hostile state, a proxy group, or independent actors.

Another defining characteristic of cyberspace operations is their relatively low deployment cost, paired with the potential for high impact, making them a double-edged sword in warfare. On the positive side, low-cost strategies can yield significant results. For example, a well-executed phishing campaign or ransomware attack can disrupt critical infrastructure, steal sensitive data, or create widespread panic, all without requiring substantial financial investment. However, this characteristic also presents a significant challenge: adversaries with limited resources can leverage these same strategies to inflict considerable damage. Consequently, this dynamic increases the likelihood that specific resources or pivotal actions will tip the scales in a conflict.

Another characteristic of cyber warfare is rapid technological advances. The sophistication of cyber threats continues to escalate as adversaries increasingly leverage advanced technologies like artificial intelligence and machine learning. These tools enable attackers to improve the precision, scale, and impact of their operations, making it even more challenging for defenders to maintain effective countermeasures. In this dynamic landscape, cyber warfare resembles a perpetual cat-and-mouse game, where offensive and defensive technologies evolve in tandem. The side capable of developing the most advanced tools and technologies often gains a significant technological advantage. However, a critical factor in cyber warfare is the ability to operationalize these technologies effectively within missions. The strategic deployment and integration of technology into cyber tactics and operations are decisive in determining outcomes. Consequently, the current state of cyber warfare requires not only constant innovation in cybersecurity techniques and measures but also strategic advancements in how these tools are planned, managed, and employed.

Classical warfare theories are epitomized by Clausewitzian Theory and Sun Tzu's Principles, both of which provide foundational frameworks for understanding the art and science of conflict. Clausewitzian Theory, rooted in Carl von Clausewitz’s seminal work On War \cite{clausewitz1976onwar}, presents a comprehensive framework for understanding the dynamics of conflict. A cornerstone of this theory is the ``Remarkable Trinity,'' which highlights the interplay between the government, military, and the people, emphasizing that war is as much a political and social endeavor as a military one. Clausewitz also introduces the concept of the ``Fog of War,'' which refers to the inherent uncertainty and unpredictability in military operations. This uncertainty complicates decision-making and requires adaptability and resilience. The ``Center of Gravity'' is another key tenet, representing the critical source of an adversary's strength or focus of effort, which, when targeted, can lead to decisive outcomes. Finally, the theory addresses ``Friction,'' the myriad of unpredictable factors that disrupt plans and hinder operations, underscoring the necessity of flexible strategies and robust leadership.

Sun Tzu’s The Art of War offers a timeless collection of principles that prioritize strategy, adaptability, and psychological acumen over brute force \cite{tzu2008art}. Central to his teachings is the concept of deception, with the assertion that all warfare is fundamentally rooted in misdirection. By misleading the enemy, a commander can gain a decisive advantage. Equally important is the emphasis on knowledge of oneself and one’s adversary. Sun Tzu famously states that understanding both is essential to victory in a hundred battles. Additionally, the principles highlight the importance of speed and adaptability, advocating for swift, decisive actions and the flexibility to adjust tactics as conditions evolve. These ideas underscore the relevance of strategy as a means of subduing the enemy without direct conflict, emphasizing efficiency and calculated precision over prolonged engagements.

As warfare evolves, modern theories increasingly focus on system-level dynamics and the integration of advanced technology. A critical concept in this paradigm is information dominance, which emphasizes the importance of gaining and maintaining superior knowledge of the operational environment. This dominance enables more effective decision-making, precise targeting, and a reduction in the ``fog of war''. Using real-time intelligence and network communication systems, combat effectiveness is significantly improved \cite{8599853,10935636}.


\subsection{The Need for a Theory of Operations in Cyber Warfare}

Despite the growing prominence of cyber warfare as a critical domain of conflict, there remains a conspicuous lack of a unified theory to underpin operations in this field. Operational aspects are often overlooked, as much of the existing research focuses on tactical and technical levels, such as improving specific technologies for attacks or defenses. For example, significant attention is paid to techniques such as honeypot configurations, intrusion detection methods, and the application of large language models (LLM) to enhance detection capabilities \cite{yang2025largelanguagemodelsnetwork,zhu2025llmnash,zhu2025llmstack,zhu2025gamellm}. However, less emphasis is placed on operational-level considerations, namely, how to effectively utilize these techniques and technologies to achieve successful outcomes in cyber operations.

One major challenge is the literature's predominant focus on specific incidents or narrow technical aspects. Developing an operational theory requires a holistic perspective and a systematic understanding of the principles, objectives, and dynamics underlying cyber conflict. In addition, cyber warfare differs fundamentally from traditional forms of conflict. It operates in an intangible, global, and decentralized domain involving a wide array of actors, from nation-states to non-state groups. For example, attribution is difficult in the cyber domain. Attackers can mask their identities or use proxies, making retaliation and accountability highly complex. It leads to limited, often incomplete information, exacerbating decision-making challenges during cyber operations.

Another challenge of operating in cyber warfare arises from the need to integrate with traditional domains such as land and air warfare. It is especially critical in multi-domain warfare, where cyber and other domains must work together to enhance each other's operations to achieve the central high-level goal. In particular, cyber operations often blur the line between war and peace, typically occurring below the threshold of conventional armed conflict. These unique dynamics demand a theoretical framework capable of addressing critical questions such as deterrence, escalation, and the establishment of international norms specific to cyberspace.

The fast-paced evolution of technology adds another layer of complexity. An operational framework must be flexible and adaptive, accommodating emerging technologies such as artificial intelligence \cite{johnson2019artificial,hartmann2020next} and quantum computing \cite{krelina2021quantum,radanliev2025cyber}. These technologies not only enhance existing capabilities but also introduce new risks or exacerbate current vulnerabilities in the cyber domain. For example, anticipating how AI-driven systems might change attack strategies or how quantum computing could undermine cryptographic methods is essential for effective operations.

Developing a unified operational theory for cyber warfare is imperative to address these challenges. This framework must provide a systematic approach to understanding and orchestrating cyber operations while remaining adaptable to the unique and evolving characteristics of the cyber domain. Only with a comprehensive theory can decision-makers effectively align tactics, technology, and strategy to achieve their objectives in this critical area of modern conflict.

\subsection{Toward A Theory of Cyber Warfare}

The lack of a comprehensive theory for cyber warfare represents a significant gap in both academic and strategic fields. Developing such a theory is essential to understanding the unique dynamics of cyber conflict, guiding effective policy making, and preparing for future challenges. As cyber warfare becomes an increasingly central element of global security, a robust theoretical framework is no longer a luxury but a necessity.

Developing a theory of cyber warfare requires a framework that accounts for the unique characteristics of cyberspace, integrates principles from traditional warfare theories, and adapts to the rapidly evolving nature of digital conflict. A promising approach could draw on systems theory and game theory to capture the complex interactions within cyberspace, involving diverse actors, advanced technologies, and widespread vulnerabilities.

Adapting Carl von Clausewitz’s principles of warfare to the cyber domain offers a particularly compelling foundation. Clausewitz’s framework \cite{clausewitz1976onwar}, with its ``remarkable trinity'' of actors (people, military, and government), aligns conceptually with the interplay of state actors, non-state actors, and infrastructure in cyberspace. This trinity can be interpreted through the lens of system theory, emphasizing interdependencies, feedback loops, and emergent behaviors. Furthermore, the inherent uncertainties of warfare, encapsulated in Clausewitz's notion of ``fog of war'', resonate with the unpredictable and dynamic nature of cyber operations, where incomplete information and deception are prevalent.

Integrating game theory into this adapted Clausewitzian model provides tools to analyze strategic interactions among cyber actors, each with distinct objectives and varying levels of information. Game-theoretic concepts \cite{kamhoua2021game} such as Nash equilibria, signaling, and bargaining dynamics could elucidate how cyber conflicts unfold, how escalation or deterrence could be managed, and how trust or cooperation could be promoted in a domain rife with mistrust.

This framework not only improves strategic clarity but also serves as a foundational guide for effective policy and operational decision-making in cyber warfare. It operates across multiple echelons of decision-making, ranging from the policy level, where cyber deterrence strategies and technology investments are shaped, to the strategic level, where overarching cyber conflict planning takes place. At the operational level, it informs the coordination of cyber defense and offensive actions, while at the tactical level, it helps optimize specific maneuvers such as intrusion detection, counter-deception, and automated response mechanisms. 

Game-theoretic frameworks provide a structured approach to modeling adversarial interactions in cyber warfare across the policy, strategic, operational, and tactical levels. However, given the complexity and speed of modern cyber operations, human decision-making alone is not sufficient. Agent-based technologies integrated with AI techniques can alleviate this burden by automating analysis, strategy development, and real-time response mechanisms \cite{loevenich2024training}. These models simulate cyber actors, including attackers, defenders, and neutral entities, using game-theoretic principles to predict and counter adversarial behavior. The combination of game theory, AI, and agent-based modeling facilitates decision dominance, the ability to consistently outpace, outmaneuver, and out-adapt cyber adversaries. By automating key aspects of cyber warfare decision-making, these technologies enable faster, more intelligent, and more resilient cyber defenses, ensuring a proactive and adaptive security posture in the rapidly evolving threat landscape.

\subsection{Organization of the Chapter}

This chapter is organized as follows. Section 1 introduces the concept of cyber warfare, highlighting its unique characteristics and complexities, and motivates the need for a unified theoretical framework to support planning and execution in this evolving domain. Section 2 identifies the fundamental components of cyber warfare and maps them to game-theoretic primitives such as players, payoffs, and strategies, providing a foundation for systematic modeling.
Section 3 establishes a game-theoretic foundation for modern cyber warfare, showing how risks, paradoxical logic, and principles of surprise can be formalized through classical and modern game-theoretic models. 

Section 4 develops a multi-echelon game-theoretic paradigm, presenting the hierarchy of policy, strategic, operational, tactical, and technical levels, and introducing the notion of warfare equilibrium that ensures forward and backward consistency across levels.
Section 5 presents a taxonomy of cyber warfare through the lens of game theory, classifying conflicts by capabilities, actors, objectives, and methods, and linking them to appropriate game-theoretic models. Section 6 illustrates the framework through a case study of a synthetic conflict, ``RedCyber,'' between China and Taiwan, demonstrating how the multi-echelon paradigm and taxonomy can be applied in practice. Finally, Section 7 concludes the chapter, summarizing key insights and laying out directions for future research.

\section{Components of Cyber Warfare}

Cyber warfare is a multifaceted domain that integrates various elements to achieve strategic, operational, and tactical objectives. Its components can be categorized as follows.

\subsection{Actors}
Cyber warfare involves a complex ecosystem of actors, each with distinct motivations, resources, and roles. States are among the most prominent participants, leveraging cyber operations to protect national interests or disrupt those of adversaries. State-sponsored activities often aim to gather intelligence, weaken rivals, or assert dominance in geopolitical contexts. For example, the Stuxnet worm, reportedly developed by the U.S. and Israel, targeted Iran's nuclear enrichment facilities, causing physical damage and delaying their program \cite{farwell2011stuxnet}. Similarly, China's APT10 group \cite{sayegh2023spotlight} has been linked to cyber-espionage campaigns targeting critical industries worldwide, underscoring how states use cyber capabilities for both defensive and offensive purposes.

In addition to states, non-state actors play significant roles in cyber warfare. These include hacktivists, such as the Anonymous group, which conducts politically motivated cyber campaigns against governments and corporations. Terrorist organizations and independent cyber-criminal groups also operate in this domain, often with disruptive or financial goals. For example, ISIS has used cyber tools to spread propaganda and recruit members, while REvil, a cyber-criminal group, has executed ransomware attacks against corporations, demanding hefty payments to restore operations \cite{sayegh2023revil}. Non-state actors can act independently or as proxies for states, further complicating attribution and response.

Alliances, such as NATO (North Atlantic Treaty Organization), represent collective efforts to address the growing threat of cyber warfare. These entities facilitate coordination among member states for both cyber defense and offensive measures. For example, NATO's cyber defense pledge emphasizes collective responsibility in strengthening cybersecurity in its members, while NATO's cyber exercises simulate attacks to improve preparedness. Such alliances are crucial for coordinating resources, sharing intelligence, and responding effectively to transnational cyber threats.

Private organizations are both targets and participants in cyber warfare. Corporations are often attacked due to their role in critical infrastructure or for financial gain. For example, the Colonial Pipeline ransomware attack disrupted fuel supplies in the U.S. and highlighted vulnerabilities in private sector systems. However, private organizations may actively participate in cyber operations. Technology companies often collaborate with governments to counter cyber threats; for example, Microsoft has worked with global partners to disrupt botnets such as TrickBot, a network of compromised devices used for cybercrime \cite{microsoft2020trickbot}. Private cybersecurity firms also play a crucial role in identifying and mitigating threats, further blurring the lines between state and non-state efforts in cyber warfare.

These actors collectively contribute to the dynamic and multifaceted nature of cyber warfare, where traditional conflict boundaries are increasingly intertwined with digital operations. Understanding their roles and interactions is key to developing effective strategies to manage and mitigate cyber risks.

\subsection{Objectives}

The goals of cyber warfare are diverse, reflecting the multifaceted nature of cyberspace conflicts. One of the primary objectives is information theft and espionage, where adversaries seek to acquire sensitive data from governments, corporations, or individuals. For example, the 2015 breach of the US Office of Personnel Management, which exposed the personal data of millions of government employees, highlights how cyber warfare can be used to gather intelligence for strategic advantage \cite{washingtonpost2015securityclearancehack}. Espionage in cyberspace is often a precursor to more aggressive actions, providing adversaries with critical knowledge to plan attacks or influence decision-making processes.

Another significant goal is the disruption of critical infrastructure. Cyber operations targeting power grids, water supplies, or transportation systems can create widespread chaos and compromise public safety. The 2015 cyberattack on Ukraine’s power grid, which left hundreds of thousands without electricity \cite{cisa2016iralertH1605601}, illustrates how adversaries can leverage cyber capabilities to undermine a nation's operational stability. Such attacks often aim to weaken a target's resilience, demonstrating vulnerability and eroding public trust in essential services.

Cyber warfare is also used to destabilize the economy. Adversaries can disrupt financial institutions, manipulate stock markets, or launch ransomware attacks to cripple economic activities. The WannaCry ransomware attack in 2017, which affected hundreds of organizations worldwide, including hospitals, disrupted critical operations and caused financial losses \cite{ibm2022wannacry}. Such actions can destabilize national economies, weaken political systems, and provide adversaries with strategic and monetary gains.

A growing focus in cyber warfare is psychological manipulation, where adversaries exploit digital platforms to influence public opinion, spread misinformation, or incite social unrest. For example, state-sponsored misinformation campaigns during elections can manipulate voter behavior, erode trust in democratic institutions, and polarize societies. The use of bots and fake accounts to amplify divisive content on social media further demonstrates how cyber operations can achieve psychological objectives, creating societal discord without direct physical confrontation.

Ultimately, many cyber operations aim to achieve strategic superiority, using the digital domain to gain long-term geopolitical or military advantages. For example, cyberattacks can be used to degrade an adversary’s defense capabilities or gain leverage in negotiations. The Stuxnet worm, which targeted Iran’s nuclear enrichment facilities, exemplifies how cyber warfare can be strategically deployed to delay or derail critical programs of an adversary without resorting to traditional military intervention.

In sum, the goals of cyber warfare span a wide spectrum, from covert information gathering to overt disruption and psychological manipulation. These objectives are often interconnected, with one goal, such as espionage, serving as a means to achieve another, such as strategic superiority. The examples highlight the transformative power of cyber operations, which have become central to modern conflicts across multiple domains.

\subsection{Domains}
While cyberspace serves as the primary domain for cyber warfare, its effects are far-reaching, often intersecting with the physical domains of land, sea, air, and space. Cyber actions can have tangible impacts in these areas, creating real-world consequences that extend beyond digital boundaries. For instance, cyberattacks targeting the electrical grid can disrupt power supply to entire regions, affecting households, businesses, and critical services. Similarly, satellite hacking can compromise GPS systems, affecting aviation, maritime navigation, and military operations. A notable example is the alleged 2014 cyberattack on a German steel mill, where hackers accessed industrial control systems, causing significant physical damage to a blast furnace \cite{bbc2014sonyhack}. These intersections highlight the potential for cyber operations to inflict harm in the physical world, blurring the line between cyber and traditional warfare.

Critical infrastructure is another high-stakes target in cyber warfare, given its importance for social functioning and security. Power plants, water supply systems, transportation networks, and healthcare facilities are increasingly reliant on interconnected digital systems, making them vulnerable to cyberattacks. For example, the 2021 Colonial Pipeline ransomware attack disrupted fuel supplies across the southeastern United States, causing widespread panic and economic impact \cite{beerman2023review}. Similarly, attacks on healthcare networks, such as the ransomware incident affecting Ireland's health service in 2021, delayed medical treatments and jeopardized patient safety \cite{reuters2021irishransomware}. These incidents underscore the critical need for robust cybersecurity measures to protect infrastructure that underpins everyday life and national security.

The human domain is also a key intersection, as cyber warfare increasingly exploits social vulnerabilities to achieve its objectives. Psychological operations (psyops), misinformation campaigns, and social engineering tactics are used to manipulate individuals or groups \cite{yang2025human}. For example, misinformation campaigns during elections aim to influence public opinion, undermine trust in democratic institutions, or polarize societies. In the realm of social engineering, phishing attacks trick individuals into divulging sensitive information, enabling further cyber intrusions. The 2016 US presidential election saw allegations of state-sponsored misinformation campaigns on social media platforms, demonstrating how human perceptions and behaviors can be influenced through cyber means \cite{yang2025transparent}. By targeting the human domain, adversaries amplify the impact of their cyber operations, turning individuals into unwitting participants in their strategies.

These examples illustrate that while cyber warfare originates in the digital realm, its consequences resonate across physical, infrastructural, and human domains. Understanding and addressing these intersections is crucial for developing comprehensive defense strategies that account for both digital and real-world impacts.

\subsection{Resources}

Effective cyber warfare relies on access to a variety of resources, which may be centralized, controlled by a single entity, or distributed among multiple actors. These resources are crucial for planning, executing, and sustaining cyber operations, often determining the success or failure of a campaign.

Human resources form the backbone of cyber warfare, which involves skilled individuals such as hackers, strategists, and analysts. Hackers may develop custom malware to breach an adversary's systems or conduct social engineering attacks to extract sensitive information. Strategists design overarching plans that align cyber operations with broader objectives, while analysts interpret intelligence to identify vulnerabilities or predict adversary movements. For instance, during the Stuxnet attack on Iranian nuclear facilities, human expertise was pivotal in crafting malware that precisely targeted industrial control systems while avoiding detection.

Knowledge resources encompass specialized information, such as exploits for zero-day vulnerabilities or insights into system weaknesses. A zero-day vulnerability is a software flaw unknown to the software’s developer, which can be weaponized before a patch is available. For example, the EternalBlue exploit, developed using a zero-day vulnerability in Windows systems, was famously leveraged in the WannaCry ransomware attack. Such knowledge allows actors to breach systems without raising immediate alarms, often causing widespread disruption before mitigation is possible.

Physical resources play a significant role when cyber operations intersect with the tangible world. These include tools and methods for disrupting infrastructure, such as physically severing communication lines, accessing secured servers, or tampering with devices. For example, during the 2015 cyberattack on Ukraine's power grid, attackers combined digital tools with physical access to substations to orchestrate a large-scale blackout. This shows how physical interventions can amplify the effects of cyberattacks on critical infrastructure.

AI resources are increasingly integral to modern cyber warfare, enabling automated tasks that enhance operational efficiency and scalability \cite{rjoub2023survey,mohamed2025artificial}. Artificial intelligence can be used for real-time reconnaissance, identifying vulnerabilities in adversary networks more quickly than human operators. It is also pivotal in crafting sophisticated misinformation campaigns, such as generating deepfake videos to undermine public trust. In addition, AI-driven tools can execute cyberattacks autonomously, selecting targets or adapting strategies based on changing conditions. For example, AI-powered bots might be deployed in a DDoS attack, overwhelming targeted servers with traffic while dynamically adjusting patterns to evade countermeasures.

By combining these diverse resources, cyber warfare campaigns can achieve precision, scalability, and impact, demonstrating the importance of aligning human expertise, technical knowledge, physical tools, and AI-driven automation in a cohesive strategy.

\subsection{Operational Elements}

Cyber warfare functions across multiple levels, each serving a distinct purpose in achieving overarching strategic objectives. These levels range from high-level planning to the execution of precise tactics, reflecting the complexity and multifaceted nature of cyber conflicts.

At the highest level, strategy encompasses comprehensive planning that guides the sequencing of operations to accomplish long-term objectives. For example, a nation-state may strategically prioritize disabling an adversary’s critical infrastructure during a conflict to create chaos and weaken its operational capabilities. This type of strategic planning often involves identifying key targets, allocating resources, and coordinating multiple campaigns to maximize impact and efficiency.

Operations are practical implementations of strategies that encompass both defensive and offensive efforts in cyber warfare. Defensive operations focus on protecting systems by mitigating vulnerabilities, detecting threats, and neutralizing attacks. For instance, a government agency might deploy advanced intrusion detection systems to thwart cyber intrusions targeting sensitive data. These measures are essential to maintain the integrity and security of critical infrastructure. In contrast, offensive operations involve proactive measures aimed at disrupting or disabling adversary systems. An example of this would be launching a cyberattack against an enemy's command-and-control servers to disrupt their military operations. Such actions are designed to impair the adversary's capabilities and gain a strategic advantage.

To effectively understand and execute these operations, the kill chain concept has been introduced. Similar to traditional military operations, the kill chain in cyberspace involves understanding how adversaries plan and execute cyber attacks. This framework can be used to capture and model offensive operations, allowing defenders to anticipate and counteract potential threats. Similarly, a defense chain can be utilized to describe defensive operations. This concept outlines the steps taken to protect systems from cyber threats, emphasizing the importance of a structured approach to defense in the ever-evolving landscape of cyber warfare. Together, these frameworks provide valuable insight into offensive and defensive strategies in cyberspace.

Cyber operations can create psychological impacts on adversaries by undermining their confidence in their information systems and decision-making processes. This aspect emphasizes the importance of information warfare as part of cyber strategy.
Cyber attacks can be classified as opportunistic or targeted. Opportunistic attacks are often automated and less focused, whereas targeted attacks aim at specific vulnerabilities within an adversary's key cyber terrain.

Within operational frameworks, information operations play a critical role. These involve tactics such as spreading misinformation, employing deception, and conducting psychological operations (psyops) to manipulate perceptions and gain tactical advantages \cite{10685605}. For example, during a geopolitical crisis, a state might disseminate false information to create confusion and delay the adversary’s response, setting the stage for a decisive offensive cyber operation.

Attacks represent the specific engagements within cyber warfare, akin to battles in traditional conflict. These can take various forms. For example, Advanced Persistent Threats (APTs) are a class of sophisticated, stealthy intrusions that persist over extended periods, such as an APT group infiltrating a financial system to siphon funds or gather intelligence \cite{alshamrani2019survey}. DDoS Attacks can overwhelm a network with traffic to render it unusable, often employed to disrupt critical services like online banking or government portals. In addition, Jamming Communications blocks or interferes with communication signals, potentially crippling military coordination or emergency response systems.

Finally, tactics involve the specific methods and techniques used to execute these attacks. These are often guided by established frameworks like MITRE ATT\&CK \cite{strom2018mitre}, which provides detailed documentation of adversarial Techniques, Tactics, and Procedures (TTPs). For example, a cyber-criminal group might use a phishing campaign to deliver malware (technique), followed by privilege escalation (tactic), to gain control over critical systems. By studying and leveraging frameworks such as MITRE ATT\&CK, defenders can better anticipate and counter these tactics, enhancing their resilience against evolving threats.

Together, these levels of cyber warfare form a cohesive system that enables actors to conduct complex and multifaceted campaigns, whether to defend against threats or to achieve offensive objectives. The interplay between these levels underscores the sophistication and high stakes of modern cyber conflicts.

\subsection{External Factors}

Cyber operations are significantly shaped by external considerations, which establish the boundaries and constraints within which cyber actors operate. Among these factors, compliance with legal and regulatory frameworks is paramount. Nations, organizations, and alliances must adhere to international treaties, national laws, and organizational policies that govern the conduct of cyber operations. For example, the Tallinn Manual provides guidelines on the applicability of international law to cyber warfare, defining acceptable conduct during conflicts \cite{schmitt2017tallinn}. Similarly, organizations must comply with standards such as the General Data Protection Regulation (GDPR) in the European Union, which governs the handling of personal data and restricts certain cyber activities. Non-compliance can result in legal penalties, diplomatic fallout, or reputational damage, emphasizing the need for operations that align with established norms.

Constraints also play a critical role in shaping cyber operations, encompassing technical, geopolitical, and resource limitations. Technological constraints, such as the availability of advanced tools or expertise, may restrict the scope and sophistication of an operation. For instance, a state with limited cybersecurity infrastructure might struggle to defend against or execute APTs. Geopolitical factors, such as alliances or rivalries, influence the planning and execution of cyber campaigns. A nation might hesitate to launch a cyberattack if it risks escalating tensions with powerful adversaries or alienating allies. Resource limitations, including financial constraints or human resource shortages, can further hinder operations. An example is a small nation relying on third-party contractors for cybersecurity defense, which may leave it vulnerable to more sophisticated state-sponsored attackers.

Finally, ethics introduces an important dimension to cyber operations, dictating what actions are considered acceptable based on societal norms or organizational principles. Ethical considerations often address questions such as whether deception, misinformation, or harm to civilian systems is permissible. For instance, targeting a hospital's network during a conflict would be deemed unethical and potentially a violation of international humanitarian law, even if the attack disrupts an adversary's operations. On the other hand, cyber deception \cite{Zhu2025CyberDeception}, such as deploying honeypots to mislead attackers, is often considered an ethical and effective defensive measure. However, ethical norms can vary widely across cultures and organizations, leading to differing interpretations of acceptable conduct. For example, some nations may see cyber espionage as a legitimate extension of traditional intelligence-gathering, while others view it as a violation of sovereignty.

\subsection{Outcomes}

The outcomes of cyber warfare are multifaceted and can be evaluated at various levels, depending on the objectives and the broader context of the conflict. At the most basic level, operations are assessed as wins, losses, or neutrals, reflecting their immediate effectiveness. For example, a successful DDoS attack that temporarily disrupts an adversary's critical infrastructure might be considered a ``win'' in the context of the operation, achieving its intended goal of creating disruption. Conversely, if a cyber defense system thwarts such an attack without significant impact, it would count as a ``loss'' for the attacker. Neutral outcomes occur when the operation fails to produce meaningful effects or when the benefits and costs balance out, leaving neither party significantly advantaged.

However, the outcomes of cyber warfare extend beyond isolated engagements, encompassing multilevel outcomes that reflect the interplay between individual operations and overarching strategic goals. In this context, success or failure in a specific operation, a ``battle'', does not necessarily determine the success or failure of the larger campaign, the ``war.'' For instance, an APT operation might fail to infiltrate a high-value target due to robust defenses. Although this may seem like a loss, the reconnaissance conducted during the attempt could yield valuable intelligence that enables future operations to succeed, contributing to a longer-term strategic objective.

A notable example of multilevel outcomes is the Stuxnet attack, where a highly sophisticated cyber operation targeted Iran's nuclear enrichment facilities. While the operation succeeded in temporarily disabling centrifuges, it also revealed vulnerabilities in industrial control systems that adversaries have since sought to exploit globally. For the attackers, the operation was a tactical success but also a strategic tradeoff, as the knowledge gained by defenders and subsequent tightening of cybersecurity around critical infrastructure complicated future efforts.

It is important to recognize that outcomes in cyber warfare are not always zero-sum. Both parties can suffer losses, and the consequences can extend beyond the primary actors, resulting in significant collateral damage.
Cyber warfare can be characterized using a framework defined by key elements. Each element corresponds to a game-theoretic counterpart. Clearly defining these elements facilitates the formulation of a game and provides valuable insights for its analysis, which will be explored in the next section.

\section{Establishing a Game-Theoretic Foundation for Modern Cyber Warfare}

Cyber warfare is marked by rapid technological change, asymmetry of power, and the interplay of technical, organizational, and societal layers. Traditional static models or purely technical defenses cannot capture this adaptive and adversarial landscape. Game theory provides a principled foundation by treating attackers, defenders, and intermediaries as decision-makers with conflicting objectives. It formalizes incentives, equilibria, and strategic trade-offs, including direct confrontation, signaling, deception, coordination, and escalation. In doing so, it offers a unifying language to analyze resilience, deterrence, and resource allocation across multiple levels of conflict. This section develops that foundation, preparing the ground for a structured analysis of multi-echelon operations, equilibrium concepts, and case studies that illustrate the strategic logic of cyber conflict.

\subsection{Multi-Echelon Warfare}
Cyber warfare can be analyzed and executed at multiple levels: policy, strategy, operations, and tactics. Each level plays a critical role in shaping the overarching objectives of a cyber campaign, with each stage building upon the foundations laid by the preceding one.

\subsubsection{Policy Level}
At the policy level, cyber warfare is linked to \emph{grand strategy}, also known as high strategy, which extends beyond the immediate concerns of military and cyber operations. Grand strategy integrates economic relations, diplomatic behavior, and international collaboration to achieve long-term objectives. It involves leveraging all available instruments of national power, diplomatic, informational, military, and economic (the DIME framework) to ensure the security and prosperity of a nation or coalition. This level addresses the nation-state scale, where policies shape the operational and tactical choices that follow.

\subsubsection{Strategy Level}
At the strategy level, the focus is on high-level planning to align operations with the overarching objectives of warfare. This involves developing a sequence of defensive and offensive operations, allocating resources to maximize effectiveness, and ensuring coordination between interdependent operations. For example, if the objective is to disrupt an adversary’s critical infrastructure, the strategy could involve a combination of information operations to create misinformation and psychological pressure, followed by a cyber offensive targeting the operating technology systems of a power grid. In this context, information operations, such as the spread of false information about the grid's stability, could create confusion and panic, thereby softening defenses and aiding the subsequent offensive operation. Resource allocation plays a pivotal role at this level, ensuring that skilled personnel, zero-day vulnerabilities, and AI-powered tools are optimally deployed in planned operations.

\subsubsection{Operation Level}
At the operational level, the focus narrows to planning specific operations that align with strategic objectives. This involves designing a sequence of attacks that, when executed, achieve the intended outcome of an operation. For instance, in an offensive operation aimed at disabling a transportation system, the plan might involve multiple steps: first, a phishing campaign to gain initial access to the transportation agency's network; next, a ransomware attack to encrypt critical operational data and disable system functionality; and finally, a DoS attack to prevent recovery efforts. These attacks are not isolated; they are composed so that each amplifies the impact of the others, ensuring the operation achieves its intended goal. The success of an operation depends on the ability to identify vulnerabilities, sequence attacks effectively, and adapt to the adversary’s countermeasures.

\subsubsection{Tactical Level}
The tactic level focuses on the detailed execution of individual attacks (defenses), determining the specific techniques, tactics, and procedures (TTPs) required to carry them out successfully. For example, in a phishing attack, tactics might include crafting highly personalized emails that mimic trusted sources and leveraging psychological manipulation to increase the likelihood that the target will click malicious links. In a DoS attack, tactics might involve deploying botnets to overwhelm servers with traffic while simultaneously masking the source of the attack to delay mitigation. The MITRE ATT\&CK framework is often used at this level to plan and refine tactics, ensuring that each attack is optimized for maximum impact.

\subsubsection{Technical Level}
The associated technical level provides the technological foundation to enable and enhance these tactics. Technological advances, particularly in areas like AI, automation, and encryption, are reshaping the landscape of conflict, enabling more sophisticated and impactful TTPs. Together, these levels work in tandem, with tactical strategies that guide the use of cutting-edge technologies and technical innovations that expand the possibilities for tactical execution. This interplay underscores the importance of staying ahead in both strategy and technology to succeed in modern conflict environments.

\subsubsection{Interdependencies Between Levels}
The three levels of cyber warfare are highly interconnected. Strategic planning provides the overarching framework that dictates the goals and resource allocation for operations. The operational level translates these strategic goals into actionable plans composed of multiple coordinated attacks (defenses). The tactical level ensures that each individual attack (defense) within an operation is meticulously planned and executed to effectively contribute to the overall success. For example, a campaign to disrupt an adversary's communication infrastructure might strategically plan to exploit interdependencies between networks and users, operationally design a series of cascading attacks on routers and satellites, and tactically deploy precision malware to exploit specific vulnerabilities in routers.

By addressing cyber warfare at these three levels, attackers and defenders alike can structure their efforts to maximize impact, adapt dynamically to evolving scenarios, and achieve long-term objectives in the contested domain of cyberspace.

\begin{table}[h!]
\centering
\small
\renewcommand{\arraystretch}{1.2}
\begin{tabularx}{\textwidth}{|p{2.8cm}|p{3.5cm}|X|}
\hline
\textbf{Warfare Element} & \textbf{Game Theory Element} & \textbf{Description} \\
\hline
Actors & Players & Decision-making entities, whether individuals or groups, involved in the interaction. \\
\hline
Objectives & Utility Functions or Payoffs & Goals or desired outcomes, represented by utility or payoff functions in game theory. \\
\hline
Domains & Game Type or Mode & Context or setting of the interaction (e.g., cyber, land, sea), akin to the type or mode of the game. \\
\hline
Resources & Set of Feasible Actions or Strategy Sets & Resources define the possible actions or strategies available to the players. \\
\hline
Operational Elements & Strategy Profiles & Operational plans in warfare correspond to strategies or combinations of strategies in game theory. \\
\hline
External Factors & Environment or Exogenous Parameters & Factors outside the players' control that influence the game's dynamics. \\
\hline
Outcomes & Resulting Payoff or Equilibrium Outcome & The results of the interaction, represented by payoffs, equilibria, or terminal states. \\
\hline
\end{tabularx}
\caption{Illustration of the correspondence between warfare elements and game theory elements.}
\label{tab:warfare-game-theory}
\end{table}

\subsection{Role of Game Theory in Cyber Warfare}

\subsubsection{Understanding Risks in Cyber Warfare Through Game Theory}

Risks in cyber warfare arise from adversarial interactions, human and organizational weaknesses, systemic dependencies, and vulnerabilities at multiple levels of strategy. Game theory provides a rigorous framework for modeling these risks by analyzing the strategic interplay among attackers, defenders, and third parties within broader environments \cite{yang2024prada}. In the following, we outline key categories of risks and their game-theoretic implications.

\para{Risks from Adversarial Interactions} 
Attackers exploit vulnerabilities and disrupt systems, while defenders deploy countermeasures to contain threats. However, defensive actions carry their own risks: poorly designed responses can escalate conflict, produce unintended consequences, or cause collateral damage to civilian infrastructure or neutral actors. Game-theoretic models capture these dynamics as contests of adaptation. For example, \emph{Stackelberg security games}, \cite{an2017stackelberg,yang2025deceive} illustrate how commitment to patrols or monitoring schedules can deter adversaries, but overly aggressive commitments can backfire if attackers adapt. Similarly, \emph{signaling games} \cite{li2023price,pawlick2018modeling} explain how misinterpreted defensive signals can escalate tensions or induce attackers to test hidden capabilities.

\para{Risks from Human and Organizational Factors} 
Weaknesses within organizations introduce exploitable openings. Human errors, such as misconfigurations, delayed responses, or insider threats, intersect with resource constraints such as underfunded budgets or inadequate training. Coordination failures among technical, legal, and executive teams compound risks. Game theory highlights how internal misalignment shapes external vulnerabilities. \emph{Principal-agent models} \cite{zhang2021equilibrium,chen2018linear} capture misaligned incentives within organizations (e.g., between management and frontline responders), while \emph{Bayesian games} model how uncertainty about insider intentions leads to costly monitoring and screening \cite{huang2021duplicity,huang2019dynamic}.

\para{Risks Across Levels of Warfare} 
Risks manifest differently across the echelons. At the technical level, software bugs or unpatched systems create immediate vulnerabilities. At the tactical level, operational errors, such as mistargeting or premature exposure, undermine missions. Operational-level failures include misaligned objectives or communication overload. At the strategic level, overextension and misjudgment of adversaries compromise campaigns. At the grand strategic level, long-term risks include sanctions, political fallout, and the risk of escalation to kinetic conflict. A hierarchical game framework illustrates how local actions (e.g., disabling a power grid) propagate upward into operational disruptions, strategic instability, and national security repercussions. For instance, \emph{dynamic stochastic games} can model how a tactical intrusion cascades into operational degradation \cite{huang2021dynamic,zhu2009dynamic}, while \emph{differential games} highlight escalation dynamics over extended campaigns \cite{huang2020dynamic}.

\para{Systemic Risks and External Dependencies} 
Global interdependence magnifies risks. Third-party software and hardware introduce supply chain vulnerabilities, while geopolitical uncertainty and the behavior of neutral actors complicate planning. Reliance on shared infrastructures such as DNS systems or cloud providers creates systemic exposure, where a single failure cascades across multiple stakeholders. Game theory models these as \emph{multi-player games with externalities}, where equilibrium outcomes are inefficient unless actors coordinate. \emph{Public goods games} \cite{kosfeld2009institution,galbiati2008obligations} capture underinvestment in shared security (e.g., collective defense of Internet routing), while \emph{network games} explain how localized disruptions propagate through interdependent systems \cite{liu2023game,liu2022herd}.

\para{Game-Theoretic Insights for Risk Management} 
Game theory enables organizations to quantify risks, anticipate adversarial moves, balance trade-offs, and improve decision-making under uncertainty. Payoff structures capture expected costs and benefits, guiding the choice of dominant or mixed strategies that mitigate risks while preserving effectiveness. Importantly, game theory also provides \emph{metrics}, such as expected loss, value of the game, or the \emph{window of superiority} \cite{raio2023toward}, that quantify the consequences of different strategies and the efficiency of equilibrium outcomes. For example, \emph{Parrondo’s paradox} shows how the alternation of weak defenses individually can produce a strong overall strategy \cite{harmer2002review}, while \emph{Braess’ paradox} warns that adding defensive resources can inadvertently worsen systemic performance \cite{nagurney2021braess}. 

Anticipation and adaptation are central: \emph{honeypot games} demonstrate how deception disrupts attacker reasoning, while \emph{adaptive security games} model defenders that evolve in response to emerging threats \cite{huang2020farsighted,Zhu2025CyberDeception,pibil2012game}. Trade-off analysis directs resource allocation, for instance between detection tools and endpoint protection, under budget constraints. Finally, systematic modeling reduces the reliance on intuition when adversary capabilities or intentions remain uncertain and highlights the value of coordination across sectors to pool resources and strengthen collective defense.

\subsubsection{Understanding ``Paradoxical Logic''}

Paradoxical logic challenges the linear reasoning often applied to decision-making. In conflict, outcomes are rarely straightforward: more does not always mean better, and winning a battle does not guarantee winning a war. Such counterintuitive results arise because adversaries are intelligent, adaptive agents whose actions and reactions create nonlinear, interdependent dynamics. 

Clausewitz stated this distinction: ``War is not the action of a living force upon a lifeless mass $...$ but always the collision of two living forces''~\cite{clausewitz1976onwar}. Unlike mechanical systems with predictable responses, warfare unfolds as a strategic interaction in which each move is met by a countermove, generating complexity, unpredictability, and frequent reversals.

Linear reasoning, therefore, fails in adversarial settings. For example, preparing for every possible contingency seems prudent, but it is ultimately infeasible. The costs of covering all vulnerabilities are prohibitive, and adversarial adaptation ensures that new weaknesses will always emerge. Defensive measures thus incur dynamic costs: as defenders fortify one area, attackers redirect efforts elsewhere, driving up the overall burden of security. Strategic prioritization, focusing on critical assets and key risks, is essential to avoid diminishing returns.

Paradoxical dynamics often lead to reversals of intent, where actions produce opposite outcomes. Aggression may unite adversaries, while over-defense in one area can create exposure in another. Game theory offers tools to analyze these complexities. Several paradoxes are especially instructive:

\para{Parrondo’s Paradox.} Two losing games, such as a biased coin toss (Game A) and a capital-dependent coin toss (Game B), when alternated or randomly mixed, can form a winning strategy. This counterintuitive effect highlights how deception strategies that appear weak in isolation may, when combined, generate advantage over adaptive adversaries. In \cite{harmer1999losing}, the authors demonstrate how combining two individually losing strategies can paradoxically yield a winning outcome. Translated into cyber defense, this suggests that alternating between deceptive tactics, each of which may appear ineffective or even detrimental in isolation, can produce an overall advantage by disrupting attacker expectations and decision-making. For instance, switching between high-visibility decoys and low-profile honeypots may confuse adversaries, forcing them into suboptimal choices. This insight highlights the value of counterintuitive strategy design in cyber warfare: by embracing paradox and leveraging surprise, defenders can transform apparent weaknesses into systemic strengths.

\para{Information Paradox.} More information is not always beneficial. Akerlof’s ``Market for Lemons'' shows how asymmetric information can cause market collapse even when greater transparency is introduced \cite{akerlof1978market}. Gossner demonstrates that different information structures can strictly change equilibrium outcomes and that partial ignorance may, in some environments, sustain trade or improve incentives \cite{gossner2000comparison}. These results illustrate a deeper structural insight: increasing information does not monotonically improve performance; instead, it reshapes strategic behavior.

In cyber warfare, this paradox emerges in disclosure strategy. Defenders who reveal too much about their defensive posture may reduce uncertainty for attackers and inadvertently improve adversarial planning. At the same time, opacity can preserve uncertainty and deterrence. The tradeoff is therefore not about ``more versus less'' information, but about how information is structured and how beliefs are formed. 

Empirical evidence shows that defensive deception can remain effective even when attackers are explicitly informed that deception may be present. In the Tularosa Study, Ferguson-Walter et al. show that the presence of decoys, combined with disclosure that deception might be deployed, significantly impeded attacker progress, increased wasted effort, and altered adversarial behavior \cite{ferguson2018tularosa}. Participants who were aware of possible deception still triggered decoys at high rates and exhibited measurable decision disruption. These findings suggest that overt acknowledgment of deception does not eliminate its defensive value; rather, it can reshape attacker beliefs and incentives in ways that enhance deterrence and operational friction \cite{rogers2026tularosa}.

Recent results on the price of transparency clarify this distinction. Li and Zhu compare overt persuasion, where the signaling mechanism is publicly known, with covert signaling, where the mechanism is hidden \cite{li2023price}. They show that transparency does not reduce the sender’s equilibrium payoff relative to opacity in a perfect Bayesian equilibrium. In strictly competitive settings, including zero-sum environments, transparent and opaque structures yield the same payoff. In other classes of games, however, opacity can introduce informational friction that degrades performance, and the gap between the two regimes can become significant. Thus, contrary to intuition, transparency can be strategically advantageous rather than costly.

A similar paradox appears in multi-agent systems. Additional feedback information can change equilibrium strategies in ways that do not necessarily reduce aggregate cost. As shown in differential games with open-loop versus closed-loop structures, the efficiency comparison depends on population size and cost parameters \cite{bacsar2011prices}. Effective communication design, therefore, requires understanding how information alters equilibrium incentives, rather than assuming that more feedback automatically improves coordination.

\para{Braess’ Paradox.}
Braess’ paradox illustrates that adding capacity to a network does not always improve overall performance \cite{nagurney2021braess,di2014braess}. When users act strategically, structural improvements can alter equilibrium behavior in ways that increase congestion rather than reduce it. In some cases, removing a link can improve system-wide outcomes. 

A related observation appears in recent work on transportation resilience. The PRADA framework shows that informational perturbations, even when introduced with adversarial intent, can sometimes lead to improved equilibrium traffic conditions \cite{yang2024prada}. These findings reinforce a broader lesson: system performance depends not only on infrastructure or controls, but on how strategic agents adapt to them.

In cyber systems, the same principle applies. Adding defensive layers or controls may unintentionally introduce complexity, shift attacker incentives, or create new bottlenecks. Resilience is therefore not a monotonic function of added resources. It emerges from the interaction between structure and strategic behavior.

\medskip
These paradoxes illustrate that intuition is an unreliable guide in adversarial dynamics. Therefore, an effective cyber strategy must embrace paradoxical logic, leveraging combinations of weak tactics, calibrated information release, and counterintuitive structural adjustments to transform disadvantage into advantage.

\subsubsection{Principles of Surprise in Cyber Warfare}

Surprise is central to cyber warfare, shaping strategy and amplifying operational impact. It exploits adversarial blind spots and introduces unexpected disruptions that undermine defenses. Unlike conventional warfare, where surprise is achieved through maneuvers or speed, cyber operations leverage stealth, misinformation, and novel attack vectors. APTs exemplify this, infiltrating quietly and striking only when conditions maximize the effect. Zero-day exploits deliver shock similarly by exploiting unknown vulnerabilities before defenses exist.

Defenders, however, can claim surprise through deception. Techniques such as honeypots, honeytokens, and decoy networks create uncertainty, forcing attackers to expend effort on false targets. Randomization, i.e., dynamic IP allocation, shifting attack surfaces, and AI-driven anomaly detection, further complicates reconnaissance and prediction, reducing the adversary’s ability to plan.

Game-theoretic approaches formalize this interplay of deception and surprise. Cyber deception games, for instance, model how defenders mislead attackers about intentions and capabilities. Motivated by experiments on cognitive bias in red-team cyber range operations \cite{dib2026}, in \cite{yang2025bilevel}, cognitive triggers are designed to create microdeceptions that exploit bounded rationality, inducing adversaries into suboptimal choices by shaping their beliefs and expectations. This slows the attack process, forces attackers to waste additional effort, and ultimately creates strategic advantages for the defender.

\begin{figure}
\begin{center}
\includegraphics[width=\linewidth]{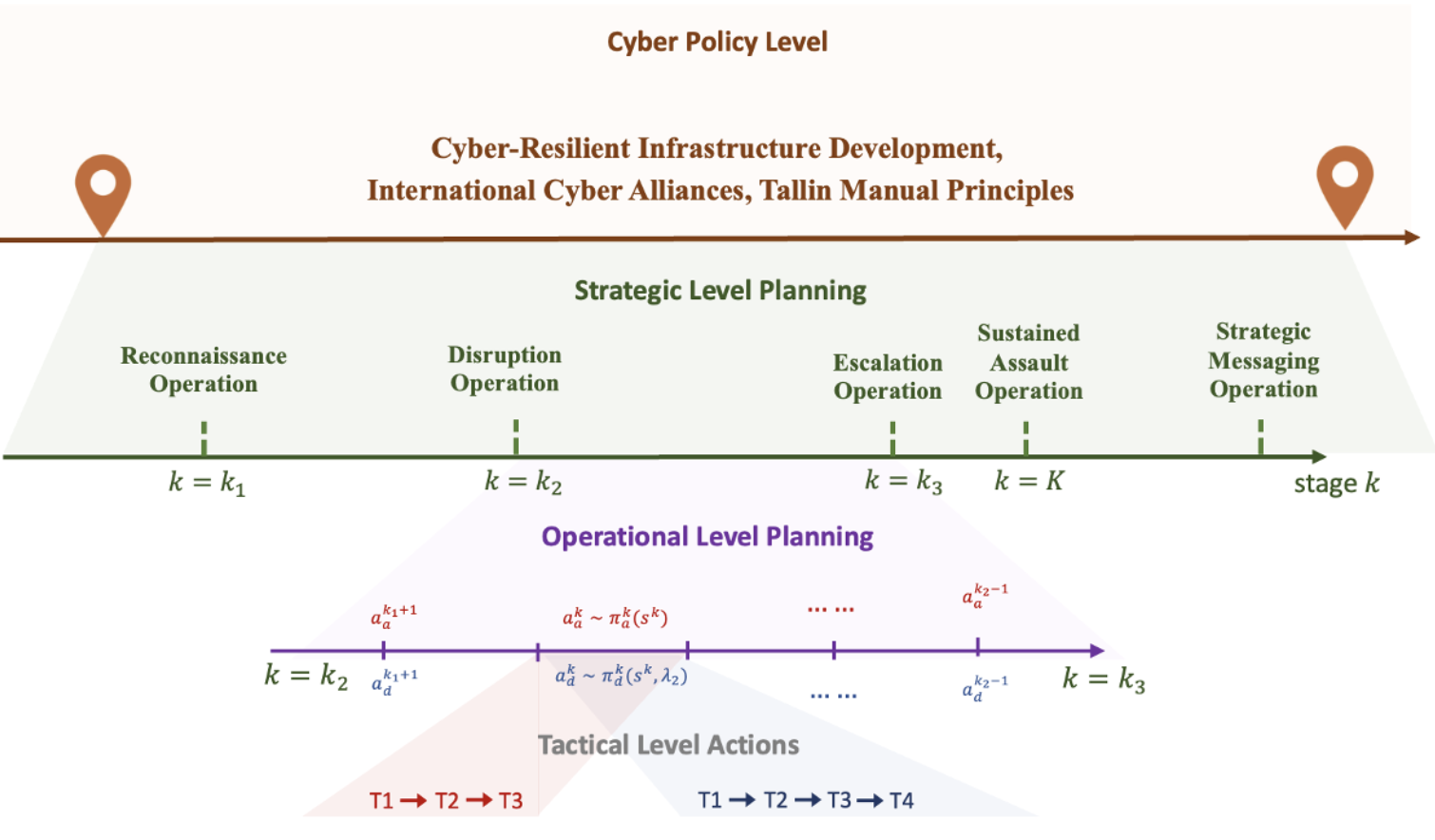}
\caption{A multi-echelon cyber warfare framework depicting the hierarchical structure of cyber conflict across policy, strategic, operational, and tactical levels. The cyber policy level defines long-term objectives through infrastructure development, alliances, and legal-normative frameworks such as the Tallinn Manual. These directives guide strategic planning, where campaigns progress through stages including reconnaissance, disruption, escalation, sustained assault, and strategic messaging. At the operational level, attacker and defender strategies evolve dynamically across stages indexed by $k$. Tactical actions (e.g., $T_1 \rightarrow T_2 \rightarrow T_3$) represent concrete maneuvers whose aggregated effects determine operational and strategic outcomes. The framework emphasizes cross-echelon interdependence: higher-level decisions shaping lower-level actions and realized outcomes feeding back to adjust strategy and policy, forming a coherent hierarchical meta-game.}
\label{fig:framework}
\vspace{-5mm}
\end{center}
\end{figure}

\section{Multi-Echelon Game-Theoretic Paradigm}

The complexity of cyber warfare arises not only from the diversity of actors, objectives, and methods, but also from the interdependence of decisions made across different levels of conflict. A purely tactical view cannot explain how budgets are set or coalitions form, while a purely strategic view overlooks the granular realities of deception, intrusion, and defense. To capture this hierarchy, we require a multi-echelon framework in which each level of warfare is modeled as a distinct but interconnected game. 

Game theory provides the mathematical rigor to represent this hierarchy. Each echelon, policy, strategic, operational, tactical, and technical, corresponds to a different game with its own players, actions, payoffs, and equilibrium concepts. 
Yet, these games are not isolated: Equilibria at one level act as constraints or objectives for the next, while outcomes at lower levels feed back to reshape higher-level decisions. This recursive dependence gives rise to what we term a \emph{meta-game}, in which stability requires consistency across all levels \cite{ge2024mega,zhu2015game}. 

In what follows, we formalize the structure of these interconnected games. 
Section~\ref{sec:multi-ech-games} introduces the key components of each level, specifying how strategies, payoffs, and equilibria are defined. Subsequent sections analyze how equilibria propagate across levels, how perturbations in one echelon affect outcomes in others, and how this framework can be used to design resilient, cross-echelon strategies in cyber conflict.

\subsection{Multi-Echelon Games}\label{sec:multi-ech-games}
Game-theoretic models are well-suited to capture the complexity of cyber warfare because they provide a rigorous way to describe interactions between attackers, defenders, and policy makers across different layers of decision-making. Figure \ref{fig:framework} illustrates a \emph{multi-echelon paradigm}, where cyber operations evolve from high-level policy decisions to low-level tactical actions. The key idea is that decisions made at higher levels (e.g., national policy) set the rules and constraints for what can happen at lower levels (e.g., individual defenses), while the outcomes at lower levels (e.g., successful or failed attacks) feed back into higher-level strategy and policy.

\para{Policy Level.}  
At the top of the hierarchy are cyber policy decisions. These include international agreements (e.g., cyber alliances, norms of behavior, and frameworks like the Tallinn Manual) and investments in national infrastructure. Such interactions can be modeled with \emph{coalition games} \cite{aumann1974cooperative}. In this setting, the players are countries, \(N=\{1,\dots,n\}\). Any subgroup of countries \(S \subseteq N\) can form a coalition, and the coalition value function \(v(S)\) describes the benefit of cooperation (e.g., shared defense, intelligence sharing). Game-theoretic solution concepts such as the \emph{core} or the \emph{Shapley value} tell us how to fairly divide the benefits of cooperation so that alliances remain stable.

\para{Strategic Level.}  
At the strategic level, cyber campaigns unfold over multiple stages, indexed by \(k \in \{k_1, k_2, \dots, K\}\), as shown in Figure \ref{fig:framework}. Each stage corresponds to a different phase of operations: reconnaissance (\(k=k_1\)), disruption (\(k=k_2\)), escalation (\(k=k_3\)), sustained assault (\(k=K\)), and finally strategic messaging.  

The state of the system at stage \(k\) is denoted by \(s^k\), which encodes information about the network, the attacker’s position, and the defender’s readiness. At each stage, the attacker follows a strategy \(\pi^k_a(s^k)\), which maps the current state \(s^k\) into a probability distribution over possible attacker actions \(a^k_a \in A_a\). The defender uses a strategy \(\pi^k_d(s^k,\lambda)\), which also depends on the state \(s^k\) and on \(\lambda\), an index that specifies which \emph{deception mechanism} is deployed (for example, honeypots, decoy files or misinformation banners). The defender’s total payoff across the campaign can be written as:  
\[
U^S_d = \sum_{k=1}^K u^S_d(s^k, a^k_d, a^k_a, \lambda),
\]  
where \(u^S_d\) is the immediate gain or loss to the defender at stage \(k\). This models how the defender must plan ahead, taking into account not only the attacker’s likely actions but also which deception tools to activate.

\para{Operational Level.}  
At this level, we zoom in on real-time planning between the attacker and the defender during a campaign. Each stage involves simultaneous moves: the attacker chooses an action \(a^k_a \in A_a\) and the defender chooses \(a^k_d \in A_d\). The system then transitions to a new state according to  
\[
s^{k+1} \sim T(s^k, a^k_d, a^k_a, \lambda),
\]  
where \(T\) is the rule (transition kernel) that describes how actions change the system. The index \(\lambda\) again matters, because deploying different deception mechanisms changes the transition dynamics (e.g., an attacker fooled by a honeypot is moved into a different state than one who bypasses it).  

As shown in the figure, the sequence of actions over time determines whether escalation occurs at stage \(k = k_3 \). Here, equilibrium concepts such as the \(\epsilon\)-Perfect Bayesian Nash Equilibrium \cite{huang2020dynamic} are important: they model how attackers form beliefs when they are uncertain about which deception mechanism \(\lambda\) is active.

\para{Tactical Level.}  
At the tactical level, engagements unfold as sequences of short-term maneuvers in which both attacker and defender adaptively select tactics over time. Unlike the zero-sum case, payoffs are generally asymmetric, so that both parties can incur losses (e.g., collateral damage) or both can gain (e.g., through intelligence gathering). Thus, tactical interactions are better represented as a \emph{nonzero-sum dynamic game}.  

At each stage $k$, the attacker chooses an action $a_a^k \in A_a$ and the defender chooses $a_d^k \in A_d$. These immediate actions constrain the feasible sets of subsequent tactic sequences,  
\[
\Xi_d^k = \Xi_d(a_d^k, a_a^k), 
\qquad 
\Xi_a^k = \Xi_a(a_d^k, a_a^k),
\]  
which encode what sequences of tactics remain possible given the stage-$k$ encounter.  A tactical game can then be expressed as  
\[
\max_{\xi_d \in \Xi_d^k} \; U^T_d(\xi_d, \xi_a), 
\qquad 
\max_{\xi_a \in \Xi_a^k} \; U^T_a(\xi_d, \xi_a),
\]  
where $\xi_d$ and $\xi_a$ are tactic sequences and $U^T_d, U^T_a$ are cumulative payoffs for defender and attacker, respectively. The outcome is a tactical equilibrium $E_{\text{tactical}}^k$, which balances the incentives of both players under the feasible sequence constraints at stage $k$.  

Graphically, the red and blue timelines illustrate this: the attacker’s tasks may evolve through $T1 \to T2 \to T3$, while the defender’s responses may expand into $T1 \to T2 \to T3 \to T4$. This illustrates both the \emph{sequence-dependence of tactics} and the \emph{asymmetric burden} of defense, where feasible sequences for the defender often require longer or more resource-intensive commitments than those available to the attacker.

\para{Technical Level.}  
Finally, at the technical level, the focus is on developing new technological and computational solutions that directly enhance cyber defense. Large Language Models (LLMs) and agentic AI serve as key enablers, enabling adaptive responses, automated decision support, and the synthesis of vast streams of threat intelligence. For instance, LLM-driven agents can monitor logs, correlate signals across domains, and propose countermeasures in real time, while multi-agent AI frameworks enable distributed coordination among defensive components \cite{zhu2025gamellm}. In addition, adversarial learning techniques, such as Generative Adversarial Networks (GANs), support the creation of realistic honeypots and synthetic data sets that improve detection and resilience \cite{gabrys2024honeygan}. At this level, the emphasis is on computational innovation that equips defenders with tools to anticipate, deceive, and outmaneuver adversaries in highly dynamic environments.



\subsection{Cross-Echelon Relationships}  

We can formalize the multi-echelon paradigm as a hierarchy of interconnected games. Let  
\[
\mathcal{G} = \{\mathcal{G}_{\text{policy}}, \mathcal{G}_{\text{strategic}}, \mathcal{G}_{\text{operational}}, \mathcal{G}_{\text{tactical}}, \mathcal{G}_{\text{technical}}\}
\]  
denote the family of games at different levels of cyber warfare, each with its own set of equilibrium \(E_\ell\). An \emph{equilibrium} here refers to a situation in which no player at that level can improve their outcome by unilaterally changing strategy, given what others are doing. The parameter $\theta$ at the technical level represents the state of the enabling technology and influences the equilibria at all the higher levels of the hierarchy.  

At the \emph{policy level}, equilibria such as coalition stability determine how nations align, how norms are set, and, most importantly, what overall objectives and resources are committed to cyber defense. We may represent policy as producing a tuple \((w,B,\theta)\), where $w$ are the weights on strategic goals (e.g., disruption vs.\ deterrence), $B$ is the total resource budget, and $\theta$ is the prevailing technological capability.  
The associated equilibrium is denoted  
\[
E_{\text{policy}}(\theta) = (w^*(\theta),B^*(\theta)).
\]  

At the \emph{strategic level}, equilibrium analysis determines both the objectives of operations and the resources assigned to each one.  
It can be modeled as a \emph{quasi-static Blotto-like game}, where players allocate finite resources across a discrete set of operations $\lambda \in \Lambda$ (e.g., deception, disruption, surveillance, or deterrence activities).  

The policy level produces outcomes $(w^*, B^*)$, where $B^*$ is the total defense budget and $w^*$ are the weights placed on different strategic objectives (e.g. deterrence vs. disruption).  
These values constrain the defender’s total resources and guide how they are prioritized across operations.  
Formally, let $R_d = B^*$ and $R_a$ be the attacker’s available resources.  
Each player then selects an allocation vector  
\[
r_d = \{r_d(\lambda): \lambda \in \Lambda, \ \sum_{\lambda \in \Lambda} r_d(\lambda) \leq R_d\}, 
\qquad 
r_a = \{r_a(\lambda): \lambda \in \Lambda, \ \sum_{\lambda \in \Lambda} r_a(\lambda) \leq R_a\}.
\]  

The effectiveness of each operation depends on the relative allocation of resources by attacker and defender, weighted by $w^*$.  
For example, the defender's payoff at the strategic level can take the form of  
\[
U_d^S(r_d,r_a;w^*,\theta) = \sum_{\lambda \in \Lambda} w^*(\lambda)\, f_\lambda(r_d(\lambda), r_a(\lambda);\theta),
\]  
where $f_\lambda(\cdot)$ captures the contribution of operation $\lambda$ to the overall success of the operation, parameterized by the technological state $\theta$.  
Similarly, the attacker’s payoff is $U_a^S(r_d,r_a;w^*,\theta)$.  
In symmetric or zero-sum cases, $U_a^S = -U_d^S$, while in nonzero-sum cases, both may incur shared costs or collateral effects.  

A canonical example of this setup is the Colonel Blotto game \cite{roberson2006colonel,hart2008discrete}, in which two players simultaneously distribute forces across multiple battlefields, and outcomes at each battlefield are determined by relative strength.  
Here, the ``battlefields'' correspond to operational categories $\Lambda$, and the policy-derived weights $w^*$ adjust the relative importance of each category.  
Equilibrium allocations determine which operations dominate, are neutralized, or remain contested.  

The associated equilibrium is denoted by  
\[
E_{\text{strategic}}(\theta) = (r_d^*(\theta;w^*,B^*), \; r_a^*(\theta)),
\]  
which fixes the allocation of resources across operations under the policy-imposed totals $B^*$ and weights $w^*$, thereby establishing the high-level objectives that constrain subsequent operational play.  

At the \emph{operational level}, a campaign is broken into phases $k=1,\dots,K$, each with a state $s^k$.  
The attacker chooses actions $a_a^k$ with strategy $\pi_a^k(s^k)$, while the defender chooses actions $a_d^k$ with strategy $\pi_d^k(s^k,\lambda)$, where $\lambda$ indexes which deception mechanism (e.g., honeypot, decoy) is deployed \cite{gabrys2024honeygan,huang2019adaptive,huang2020farsighted,yang2025deceive}.  
The defender’s cumulative payoff is  
\[
U_d^O = \sum_{k=1}^K u_d^O(s^k,a_d^k,a_a^k,\lambda,\theta),
\]  
subject to budget constraints $\sum_{k=1}^K r_k \leq r_d(\lambda)$.  
The associated equilibrium is denoted by  
\[
E_{\text{operational}}(\theta) = \big(\pi_d^{1:K,*}(\theta), \pi_a^{1:K,*}(\theta)\big),
\]  
which captures the allocation of strategies across phases.  

Dynamic equilibria such as Perfect Bayesian Nash Equilibria describe the real-time interaction of attackers and defenders under uncertainty.  
State transitions follow  
\[
s^{k+1} \sim T(s^k,a_d^k,a_a^k;\lambda,\theta),
\]  
where $T$ is the transition function and $\theta$ represents the current state of technology.  
In this dynamic view, the associated equilibrium is  
\[
E_{\text{operational}}(\theta) = \big(\pi_d^*(s^k,\lambda,\theta), \pi_a^*(s^k,\theta)\big),
\]  
which describes how attackers and defenders adapt step by step within the operation.

At the \emph{tactical level}, engagements are modeled as short-horizon contests in which both attacker and defender choose feasible sequences of tactics.  
Let $\Xi_d^k = \Xi_d(a_d^k,a_a^k)$ and $\Xi_a^k = \Xi_a(a_d^k,a_a^k)$ denote the feasible tactic sequence sets of the defender and attacker at stage $k$, determined by immediate actions $(a_d^k,a_a^k)$. Tactical interactions are naturally \emph{nonzero-sum}. An equilibrium of the stage-$k$ tactical game is defined by the pair $(\xi_d^*,\xi_a^*) \in \Xi_d^k \times \Xi_a^k$ such that  
\[
\xi_d^* \in \arg\max_{\xi_d \in \Xi_d^k} U_d^T(\xi_d,\xi_a^*;\theta),
\qquad
\xi_a^* \in \arg\max_{\xi_a \in \Xi_a^k} U_a^T(\xi_d^*,\xi_a;\theta).
\]  
The associated equilibrium is denoted by  
\[
E_{\text{tactical}}(\theta) = \big(\xi_d^*(\theta), \xi_a^*(\theta)\big),
\]  
and directly shapes operational effectiveness: strong equilibria delay attackers and expand the defender’s flexibility, while weak ones restrict operational freedom.  

At the \emph{technical level}, $\theta$ captures the state of enabling technologies (e.g., advanced LLMs \cite{touvron2023llama2,gemini2023} or GAN-based deception systems \cite{goodfellow2014generative,tolosana2020deepfakes,zellers2019defending}).  
Improvements at this level reduce costs $c_d(\theta)$, expand feasible action sets $A_d(\theta)$, alter the equilibrium strategies $(\xi_d^*(\theta),\xi_a^*(\theta))$ and their payoffs $U_d^T, U_a^T$, and change the transition dynamics $T(\cdot;\theta)$.  
The associated equilibrium is denoted \( E_{\text{technical}} = \theta^*, \) representing the dominant technological state that parameterizes the equilibria at all higher levels.  
In this way, advances at the technical level propagate upward, shifting equilibria across every higher echelon of the hierarchy.  

\para{Interdependencies.}  
The key point is that the equilibria are not isolated but are tightly linked between layers.  
At the top, policy-level equilibria $E_{\text{policy}}(\theta)$ produce outcomes $(w^*,B^*)$ that set objectives and resource budgets for the strategic game.  
Strategic equilibria $E_{\text{strategic}}(\theta)$ then allocate these resources across phases, producing constraints and feasible options for the operational level.  
Operational equilibria $E_{\text{operational}}(\theta)$ determine which tactical engagements are feasible, while tactical equilibria $E_{\text{tactical}}(\theta)$ feed into the effectiveness and costs of operations.  
Finally, the technical level permeates the entire hierarchy: the equilibrium $E_{\text{technical}}=\theta^*$ reshapes all other levels by altering costs, expanding action sets, and transforming transition dynamics.  

Formally, \emph{forward dependencies} can be expressed as  
\[
E_{\text{policy}}(\theta) \;\Rightarrow\; E_{\text{strategic}}(\theta) \;\Rightarrow\; 
E_{\text{operational}}(\theta) \;\Rightarrow\; 
E_{\text{tactical}}(\theta),
\]  
where  
\[
E_{\text{policy}}(\theta) = (w^*(\theta), B^*(\theta)), 
\]  
determines the weights and budgets available at the strategic level,  
\[
E_{\text{strategic}}(\theta) = (r_d^*(\theta;w^*,B^*), \; r_a^*(\theta)),
\]  
fixes the allocation of resources across phases, and  
\[
E_{\text{operational}}(\theta) = \big(\pi_d^*(s^k,\lambda,\theta), \pi_a^*(s^k,\theta)\big),
\]  
constrains the feasible set of tactical sequences $\Xi_d^k,\Xi_a^k$ that define  
\[
E_{\text{tactical}}(\theta) = (\xi_d^*(\theta), \xi_a^*(\theta)).
\]  

There are also \emph{backward dependencies}. Tactical equilibria $E_{\text{tactical}}(\theta)$ affect the cost and effectiveness of tactics, thereby modifying operational payoff functions $U_d^O, U_a^O$.  
Operational equilibria $E_{\text{operational}}(\theta)$ determine cumulative costs and risks, changing the strategic evaluation functions $U_d^S, U_a^S$.  
Strategic outcomes $E_{\text{strategic}}(\theta)$, in turn, modify the aggregate costs and trade-offs faced at the policy level, reshaping coalition equilibria $E_{\text{policy}}(\theta)$.  
Meanwhile, the technical state $\theta$ mediates both directions: changes in $\theta$ alter action costs $c_d(\theta),c_a(\theta)$, feasible action sets $A_d(\theta),A_a(\theta)$, and transition dynamics $T(\cdot;\theta)$.  

This establishes a two-way recursive structure:  
\[
E_{\text{policy}}(\theta) \;\leftrightarrow\; E_{\text{strategic}}(\theta) \;\leftrightarrow\; 
E_{\text{operational}}(\theta) \;\leftrightarrow\; 
E_{\text{tactical}}(\theta).
\]  
The arrows indicate not just information flow but recursive dependence: each equilibrium is both a constraint on, and a consequence of, other levels.  
Thus, a slight perturbation at one end of the hierarchy inevitably propagates to the other end.  
\[
\delta E_{\text{policy}} \;\mapsto\; \delta E_{\text{strategic}} \;\mapsto\; \delta E_{\text{operational}} \;\mapsto\; \delta E_{\text{tactical}}, 
\]  
and conversely,  
\[
\delta E_{\text{tactical}} \;\mapsto\; \delta E_{\text{operational}} \;\mapsto\; \delta E_{\text{strategic}} \;\mapsto\; \delta E_{\text{policy}}.
\]  

\subsection{The Meta-Game Warfare Equilibrium}
These forward and backward adjustments capture the inherent fragility and interdependence of cyber conflict: a new deception tactic at the tactical level may escalate costs at the operational level, shift campaign priorities at the strategic level, and ultimately reshape policy coalitions at the top.  
Conversely, a high-level policy realignment can cascade downward, altering the budgetary landscape, tightening operational constraints, and restricting tactical maneuvering.  

We define a \emph{warfare equilibrium} as a configuration  
\[
E^* = \big(E_{\text{policy}}^*, E_{\text{strategic}}^*, E_{\text{operational}}^*, E_{\text{tactical}}^*, E_{\text{technical}}^*\big),
\]  
such that both \emph{forward consistency} (higher-level equilibria correctly constraining lower-level games) and \emph{backward consistency} (lower-level outcomes coherently reshaping higher-level objectives) are satisfied.  
Formally, this may be expressed as the fixed-point condition  
\[
E^* = \Phi(E^*;\theta),
\]  
where $\Phi$ is the multi-level equilibrium mapping that integrates policy, strategic, operational, tactical, and technical dependencies.  

In other words, a warfare equilibrium is stable under mutual recursion: no level can deviate without requiring re-adjustments in all others.  
Cyber conflict is therefore best understood not as a sequence of isolated games, but as an interdependent \emph{meta game}, where equilibria at all levels must remain mutually consistent \cite{chen2019control,zhu2015game,huang2020dynamic}.  
Any perturbation, whether a policy change at the top or a breakthrough in technology at the bottom, propagates through the hierarchy, transforming outcomes across every echelon.

\begin{figure}[t]
\centering
\begin{tikzpicture}[
  >=Latex,
  node distance=20mm and 10mm,
  game/.style={draw, rounded corners, align=center, minimum width=20mm, minimum height=9mm, fill=black!3, font=\small},
  tech/.style={draw, rounded corners, align=center, minimum width=27mm, minimum height=9mm, fill=blue!8, font=\small},
  fwd/.style={-Latex, thick, blue},
  bwd/.style={-Latex, thick, dashed, red},
  techlink/.style={Latex-Latex, thick},
  lbl/.style={font=\scriptsize, inner sep=1pt, fill=white, fill opacity=0.85, text opacity=1}
]

\node[game] (policy) {
  Policy Game \\[-1pt] \scriptsize $E_{\text{policy}}(\theta)$
};
\node[game, right=of policy] (strategic) {
  Strategic Game \\[-1pt] \scriptsize $E_{\text{strategic}}(\theta)$
};
\node[game, right=of strategic] (oper) {
  Operational Game \\[-1pt] \scriptsize $E_{\text{operational}}(\theta)$
};
\node[game, right=of oper] (tact) {
  Tactical Game \\[-1pt] \scriptsize $E_{\text{tactical}}(\theta)$
};

\node[tech, above=12mm of strategic, xshift=10mm] (tech) {
  Technical Level \\[-1pt] \scriptsize $E_{\text{technical}}=\theta^*$
};

\draw[fwd] (policy) -- node[lbl, above] {forward} (strategic);
\draw[fwd] (strategic) -- (oper);
\draw[fwd] (oper) -- (tact);

\draw[bwd] (strategic.south) to[bend right=-50] node[lbl, below] {backward} (policy.south);
\draw[bwd] (oper.south) to[bend right=-50] (strategic.south);
\draw[bwd] (tact.south) to[bend right=-50] (oper.south);

\foreach \X/\pos in {policy/0.45,strategic/0.5,oper/0.5,tact/0.55}{
  \draw[techlink] (tech) -- (\X)
    node[lbl, pos=\pos, sloped, above] {$\theta$};
}

\end{tikzpicture}
\caption{Recursive interdependencies among equilibria across policy, strategic, operational, and tactical levels of cyber warfare. Forward (blue, solid) arrows represent constraints and decisions propagated downward across echelons, while backward (red, dashed) arrows capture upward feedback effects. The technical equilibrium $E_{\text{technical}} = \theta^*$ influences all levels directly. A coherent warfare equilibrium emerges only when both forward constraints and backward feedback are jointly satisfied, ensuring cross-echelon consistency.}

\end{figure}

Beyond providing structural clarity, the concept of warfare equilibrium is instrumental in explaining several critical aspects of cyber conflict.  

\para{Role of Technology.}
Because $\theta$ directly enters the mapping $\Phi$, advances in technology can disrupt equilibria at every level.  
Improved detection models, faster response systems, or novel deception tools may reduce costs $c_d(\theta)$, expand action sets $A_d(\theta)$, or accelerate transitions $T(\cdot;\theta)$.  
This explains why technological breakthroughs, whether by attackers or defenders, can dramatically change the balance of cyber warfare.  

\para{Role of Information.}
Information structures and knowledge asymmetries define effective action sets and payoff functions \cite{li2022role,yuksel2024stochastic,bacsar2011prices}.  
In some cases, more information improves strategic positioning, while in others it increases costs or generates uncertainty (e.g., information overload or deception).  
Thus, the equilibrium of warfare highlights that information can be both an advantage and a liability, depending on how it reshapes $E_{\text{operational}}$ or $E_{\text{strategic}}$.  

\para{Impact of Resources.}
The recursive structure explains the nonlinear relationship between resources and advantage.  
An additional budget $B^*$ at the policy level does not automatically translate into dominance; rather, effectiveness depends on allocation strategies $r_d(\lambda)$ at the strategic level and adaptive operations downstream.  
This captures the paradoxical logic of cyber conflict: more poorly deployed resources may yield little or no advantage.  

\para{Predictive Power under Perturbations.}
Because the equilibrium is recursive, perturbations in parameters at any level, such as $\delta\theta$ at the technical level or $\delta B^*$ at the policy level, cascade through the system.  
Studying the sensitivity of $E^*$ to such perturbations enables prediction of outcomes and identification of vulnerabilities where small changes have disproportionate effects.  

\para{Resilience and Adaptation.}
Finally, the concept of warfare equilibrium provides a foundation for designing resilience \cite{li2024symbiotic,zhu2020crosslayer}.  
Quick re-adjustment to perturbations, whether through tactical adaptation, operational redundancy, or strategic reserve allocation, ensures that the equilibrium can be restored without collapse.  
This recursive adaptability is what distinguishes fragile systems from resilient ones in cyber warfare.  

\para{Cross-Echelon Strategies.}
The warfare equilibrium also provides a framework for designing cross-echelon strategies, where decisions at one level are deliberately coordinated with those at higher or lower levels.  
This cross-layer planning ensures that tactical maneuvers support operational objectives, operational campaigns reinforce strategic goals, and strategic allocations remain aligned with policy-level coalitions and constraints.  
In practice, this allows defenders to anticipate how choices at one echelon propagate through the hierarchy, and to craft strategies that exploit these interdependencies rather than being trapped by them.  

The games-in-games framework or \emph{meta-games} for the equilibrium of warfare not only formalizes the interdependence of different levels, but also provides practical insight into the disruptive role of technology, the ambiguous value of information, the nonlinear use of resources, the prediction of cascading outcomes, the design of resilient strategies, and the development of cross-echelon decision-making.

\subsection{Anamorphism and Catamorphism Across Echelons}

To establish a rigorous framework, we formalize cross-echelon interactions through \emph{anamorphic} and \emph{catamorphic} mappings. These mappings capture the intuitive notions of \emph{zoom-in} and \emph{zoom-out} operations: anamorphisms expand coarse-grained decisions at a higher echelon into finer-grained actions at a lower echelon, while catamorphisms aggregate lower-level outcomes back into higher-level evaluations. Beyond their descriptive role, these mappings can be composed into \emph{hylomorphisms}, which ensure consistency of the equilibrium across levels by requiring that the forward unfolding and the backward folding stabilize at a fixed point. This abstraction not only parallels constructions in category theory 
\cite{awodey2010category} and program semantics \cite{meijer1991functional} but also provides an analytical tool to reason about multi-level games and cyber-physical decision hierarchies \cite{ge2024mega,zhu2015game,chen2019control}. 

\subsubsection{Mappings Between the Operational and Tactical Levels}

We can interpret \emph{anamorphism} as the process of \emph{expansion} or \emph{zooming in} within the multi-echelon paradigm. At the operational level, an action profile $(a_d^k, a_a^k)$, determined by the defender and attacker respectively, can be \emph{unfolded} into a sequence of feasible tactical maneuvers. Formally, the mapping
\[
(a_d^k, a_a^k) \;\mapsto\; (\xi_d, \xi_a) \in \Xi_d(a_d^k, a_a^k) \times \Xi_a(a_d^k, a_a^k)
\]
constitutes an anamorphism from operational-level actions into tactical-level sequences. This captures the refinement of a high-level plan into executable tactics, consistent with the notion of unfolding.

At the tactical echelon, the resulting nonzero-sum dynamic game admits an equilibrium solution $(\xi_d^*, \xi_a^*)$, which induces tactical-level equilibrium payoffs
\[
U_d^{T*}(a_d^k, a_a^k), 
\qquad 
U_a^{T*}(a_d^k, a_a^k).
\]
These represent the best-effort equilibrium outcomes conditional on the operational choices $(a_d^k, a_a^k)$.

The process of aggregating these tactical outcomes back into the operational framework corresponds to \emph{catamorphism}, or \emph{zooming out}. Specifically, the defender’s operational-level objective function can be expressed as
\[
u_d^O(s^k, a_d^k, a_a^k; \lambda, \theta) 
\;=\; U_d^{T*}(a_d^k, a_a^k) + u_d^O(s^k) + u_d^O(\lambda, \theta),
\]
where the first term reflects tactical equilibrium outcomes, the second term captures state-dependent contributions, and the third term encodes exogenous factors such as deception portfolios $\lambda$ and enabling technologies $\theta$. Thus, the upward mapping from tactics to operations is formalized as a catamorphism, collapsing detailed tactical interactions into an aggregated operational payoff.

More formally, we can define the mappings between the operational ($O$) and tactical ($T$) levels as follows.

\para{Anamorphism}
An \emph{anamorphism} is a mapping from an operational action profile to the corresponding set of feasible tactical sequences:
\[
\mathsf{ana}_{O \to T}:\ 
A_d \times A_a 
\;\longrightarrow\; 
\Xi_d \times \Xi_a,
\qquad
(a_d^k, a_a^k) \;\mapsto\; (\xi_d, \xi_a),
\]
where $A_d$ and $A_a$ denote the defender’s and attacker’s operational action sets, while $\Xi_d$ and $\Xi_a$ denote their feasible sets of tactical sequences. This mapping represents the \emph{zoom-in} process that refines a coarse operational choice into detailed tactical maneuvers.

\para{Catamorphism}
Conversely, a \emph{catamorphism} is a mapping from the equilibrium outcomes of the tactical game back to the operational objective function:
\[
\mathsf{cata}_{T \to O}:\ 
\Xi_d \times \Xi_a
\;\longrightarrow\;
\mathbb{R}^2,
\qquad
(\xi_d^\ast,\xi_a^\ast) 
\;\mapsto\; 
\big(U_d^{T\ast}(a_d^k,a_a^k),\; U_a^{T\ast}(a_d^k,a_a^k)\big),
\]
where $(\xi_d^\ast,\xi_a^\ast)$ is a tactical equilibrium and $U_d^{T\ast},U_a^{T\ast}$ are the corresponding equilibrium payoffs. These values are then aggregated into the operational-level utility, e.g.,
\[
u_d^O(s^k,a_d^k,a_a^k;\lambda,\theta)
= U_d^{T\ast}(a_d^k,a_a^k) + u_d^O(s^k) + u_d^O(\lambda,\theta).
\]

\medskip
Hence, the anamorphism $\mathsf{ana}_{O \to T}$ \emph{unfolds} operational actions into tactical sequences (zoom-in), while the catamorphism $\mathsf{cata}_{T \to O}$ \emph{folds} tactical equilibria back into operational objectives (zoom-out).

\subsubsection{Mappings Between the Strategic and Operational Levels}

Analogous to the operational-tactical case, we can define mappings between the strategic ($S$) and operational ($O$) levels.

\para{Anamorphism}
At the strategic level, players allocate resources $r_d$ and $r_a$ across a set of operations indexed by $\lambda \in \Lambda$. This induces operational policies $(\pi_d,\pi_a)$ that specify defender and attacker action sequences at the operational level. Formally, we define the mapping
\[
\mathsf{ana}_{S \to O}:\ 
(r_d,r_a) \;\longmapsto\; (\pi_d,\pi_a),
\]
where $(r_d,r_a)$ are strategic resource allocations and $(\pi_d,\pi_a)$ are the corresponding induced operational policies.

\para{Catamorphism}
Conversely, given operational equilibrium strategies $(\pi_d^\ast,\pi_a^\ast)$, we can map the outcomes back to the strategic evaluation of each operation $\lambda$. 
Let $f_\lambda$ denote the equilibrium outcome of operation $\lambda$, defined as
\[
\mathsf{cata}_{O \to S}:\ 
(\pi_d^\ast,\pi_a^\ast) \;\longmapsto\; f_\lambda,
\]
where $f_\lambda$ aggregates the equilibrium payoffs from the operational level. 
In the simplest case,
\[
f_\lambda = U_d^{O\ast}(\lambda),
\]
where $U_d^{O\ast}(\lambda)$ is the defender’s operational equilibrium payoff for operation $\lambda$. More generally, $f_\lambda$ may incorporate both defender and attacker payoffs:
\[
f_\lambda = \big(U_d^{O\ast}(\lambda),\; U_a^{O\ast}(\lambda)\big).
\]

\medskip
Hence, the anamorphism $\mathsf{ana}_{S \to O}$ unfolds strategic allocations into operational policies (zoom-in), while the catamorphism $\mathsf{cata}_{O \to S}$ folds operational equilibria back into strategic-level evaluations (zoom-out).


\subsubsection{Mappings Between the Policy and Strategic Levels}

Finally, we can formalize the relationship between the policy ($P$) and strategic ($S$) levels through anamorphic and catamorphic mappings.

\para{Anamorphism}
At the policy level, decisions determine the weight vector $w$ assigned to different operational categories $\lambda \in \Lambda$ and the total resource budget $B$. These policy parameters constrain the allocation of resources at the 
strategic level. Formally, we define
\[
\mathsf{ana}_{P \to S}:\ 
(w,B) \;\longmapsto\; (r_d,r_a),
\]
where $(r_d,r_a)$ are the defender’s and attacker’s resource allocations across operations $\lambda$, subject to the policy-level constraints.

\para{Catamorphism}
Conversely, the strategic-level equilibrium allocation $(r_d^\ast,r_a^\ast)$ leads to equilibrium payoffs $(U_d^{S\ast},U_a^{S\ast})$. These outcomes can be mapped 
back to the policy level as evaluations of the value of acting independently (i.e., without coalition support). Let $v$ denote the value of the characteristic function under one single player. Then
\[
\mathsf{cata}_{S \to P}:\ 
(r_d^\ast,r_a^\ast) \;\longmapsto\; v,
\]
where in the simplest case,
\[
v = U_d^{S\ast}(r_d^\ast,r_a^\ast).
\]

\medskip
Hence, the anamorphism $\mathsf{ana}_{P \to S}$ unfolds policy-level decisions $(w,B)$ into strategic resource allocations (zoom-in), while the catamorphism $\mathsf{cata}_{S \to P}$ folds strategic equilibrium outcomes back into the policy-level characteristic function $v$ (zoom-out).

\subsubsection{Hylomorphism Across the Echelons of Warfare}

The \emph{hylomorphism from policy to tactics and back to policy} is defined as the composition
\[
\mathsf{hylo}_{P \leftrightarrow T}
\;\triangleq\;
\mathsf{cata}_{S \to P}
\circ \mathsf{cata}_{O \to S}
\circ \mathsf{cata}_{T \to O}
\circ \mathsf{ana}_{O \to T}
\circ \mathsf{ana}_{S \to O}
\circ \mathsf{ana}_{P \to S}:\; \mathcal{P} \longrightarrow \mathcal{P}.
\]

This mapping captures the complete round trip across the four echelons. 
A policy decision $(w,B)\in\mathcal{P}$ is first unfolded into strategic allocations $\mathcal{S}$, then into operational policies $\mathcal{O}$, and finally into tactical maneuvers $\mathcal{T}$ through the successive application of anamorphisms. The resulting tactical equilibria are then aggregated back up through operations and strategy into a policy-level evaluation by applying the catamorphisms in reverse order. A policy $(w^*,B^*)$ is called a \emph{hylomorphic fixed point} if
\[
(w^*,B^*) = \mathsf{hylo}_{P \leftrightarrow T}(w^*,B^*).
\]
Such a fixed point represents consistency between directives issued at the policy level and the aggregated outcomes realized at the tactical level and folded back to policy. In practice, the convergence of the synchronized adaptation algorithm corresponds to the discovery of such a hylomorphic fixed point.

\para{Other hylomorphisms.}
While the policy-tactic cycle represents the most comprehensive round trip, hylomorphic structures can also be defined at intermediate scales, linking adjacent echelons or triplets of echelons. These partial hylomorphisms serve as localized consistency checks that ensure coherence between neighboring levels:  

\begin{itemize}
    \item \emph{Strategy-Operation hylomorphism:}
    $
    \mathsf{hylo}_{S \leftrightarrow O} 
    \;\triangleq\; \mathsf{cata}_{O \to S} \circ \mathsf{ana}_{S \to O},
    $
    which requires that strategic resource allocations, once expanded into operational policies and then aggregated back, remain aligned with the intended strategic evaluations.
    \item \emph{Operation-Tactic hylomorphism:}
    $
    \mathsf{hylo}_{O \leftrightarrow T}
    \;\triangleq\; \mathsf{cata}_{T \to O} \circ \mathsf{ana}_{O \to T},
    $
    which guarantees that operational directives, when unfolded into tactical maneuvers and folded back, yield operational outcomes consistent with the original directives.
    \item \emph{Policy-Strategy hylomorphism:}
    $
    \mathsf{hylo}_{P \leftrightarrow S} 
    \;\triangleq\; \mathsf{cata}_{S \to P} \circ \mathsf{ana}_{P \to S},
    $
    which ensures that policy-level priorities (weights and budgets) are faithfully reflected in the equilibrium consequences observed at the strategic echelon.
\end{itemize}

Here, \emph{consistency} means that a state is preserved under the round-trip
mapping: if $x$ denotes a state at echelon $X$, then
\(
x = \mathsf{hylo}_{X \leftrightarrow Y}(x).
\)
In other words, unfolding the state into a lower echelon and folding it back again does not change the state. Such a fixed point indicates that decisions and outcomes are mutually coherent across the two levels, with no residual misalignment.

Each hylomorphism therefore defines a local fixed-point condition for its associated echelons. Taken together, these local consistencies form the building blocks of the global multi-echelon hylomorphism 
\(
\mathsf{hylo}_{P \leftrightarrow T},
\)
which can be viewed as the composition of the intermediate hylomorphisms. Ensuring that each local hylomorphism admits a fixed point guarantees that adaptation remains synchronized both globally across the full policy--tactic
cycle and locally between adjacent levels of warfare.

\subsection{Warfare Equilibrium as a Lens on Stability, Dominance, and Winning}

The equilibrium concept provides a unifying framework for analyzing warfare outcomes across multiple echelons. By definition, an equilibrium captures a state in which neither side can unilaterally improve its payoff given the actions of the other. Once embedded into our multi-echelon structure through anamorphisms, catamorphisms, and hylomorphisms, equilibria serve as a natural tool for addressing three fundamental notions in warfare: the \emph{stability} of conflict dynamics, the \emph{dominance} of one side over the other, and the \emph{winning} of the war. Each builds on the other, forming a progression from robustness to superiority to decisive success.

\para{Stability.}
Stability concerns whether an equilibrium, once reached, can withstand perturbations. A stable warfare equilibrium is one in which small shocks at any level of warfare, tactical, operational, strategic, or policy, when propagated forward and backward through the mappings, do not overturn the balance of outcomes. Stability thus reflects resilience: the system has settled into a pattern where adaptive moves by either side no longer dislodge the equilibrium state. Stability is the baseline requirement for any meaningful analysis of long-term warfare outcomes.

\para{Dominance.}
Dominance builds on stability by incorporating relative performance. A dominant equilibrium for the defender is one in which, across feasible adaptations by the attacker, the defender consistently secures a superior payoff. This means that the defender not only maintains stability but also tilts the equilibrium in its favor, constraining the attacker’s effective options and shaping the escalation ladder. Dominance, therefore, reflects the defender’s ability to hold a structural advantage within equilibrium, ensuring that conflict dynamics systematically benefit one side more than the other.

\para{Winning.}
Winning is the culmination of stability and dominance, defined by equilibrium conditions that align with overarching objectives. A winning equilibrium is one in which the defender’s payoff exceeds a viability threshold, the attacker’s payoff falls below its sustainability threshold, and these conditions are preserved consistently across echelons. In this sense, winning is not limited to isolated victories but corresponds to achieving a favorable, stable, and synchronized equilibrium in which tactical gains reinforce operational and strategic success, ultimately delivering policy-level objectives.

\para{Hierarchy of concepts.}
These three notions, stability, dominance, and winning, form a natural hierarchy. Stability is the minimal condition, ensuring that equilibria persist under adaptation. Dominance strengthens stability by requiring that equilibria systematically favor one side. Winning incorporates both but goes further by imposing objective-based thresholds that capture the essence of success in war.
Taken together, they illustrate how the equilibrium framework not only describes the mechanics of conflict but also provides a principled way to reason about what it means to prevail.

\para{Why losing a battle can still mean winning the war?}
It is important to emphasize that outcomes at the tactical or operational levels do not by themselves determine the outcome of the war. Within the multi-echelon framework, these levels feed into higher evaluations through the catamorphisms, but ultimate success is judged at the strategic and policy levels.  

A tactical defeat, such as the temporary loss of a system or the failure of a particular deception maneuver, may lower the defender’s payoff at that echelon,  written as $U_d^{T} < \Theta_d^{T}$. Similarly, an operational setback, such as losing control of a campaign, may also produce local disadvantage. Yet the war can still be won if the higher-level evaluations remain favorable once these outcomes are aggregated.  

This happens for several reasons. First, strategy often involves trade-offs: resources may be conserved in one theater to achieve a decisive advantage in another. Second, local setbacks can be part of deception, drawing the adversary into costly commitments that ultimately reduce their overall payoff. Finally, what defines victory is not the outcome of isolated engagements, but whether policy objectives are achieved, such as deterrence, survival, or coalition cohesion. In equilibrium terms, the defender still ``wins the war'' if the global hylomorphic mapping preserves consistency and produces a policy-level state in which the defender's payoff meets or exceeds its required threshold, while the attacker's payoff remains at or below its corresponding threshold.

Thus, losing a battle at the tactical or operational echelon does not equate to losing the war. The essential point is that war is won at the global equilibrium level, where tactical and operational outcomes are subordinated to strategic coherence and policy success.

\section{Cyber Warfare Taxonomy}

A cyber warfare taxonomy is valuable not only for classification, but also for selecting the right game-theoretic model to study each situation. Each category connects naturally to elements of a game: the \emph{objectives} of a campaign shape the payoff functions $u_i$, the \emph{actors} define the players $N$, and the \emph{capabilities and methods} determine the feasible action sets $A_i$. Equilibrium analysis $E$ then depends on how these components interact.  

From a perspective of \emph{capabilities}, asymmetric conflicts, where weaker actors exploit low-cost attacks against stronger adversaries, are well captured by \emph{attacker-defender security games}, with unbalanced action sets or unequal costs. Symmetric conflicts, where both sides have comparable capabilities, lend themselves to \emph{zero-sum matrix games}, where $u_d=-u_a$. Escalatory campaigns are naturally modeled as \emph{repeated or stochastic games}, in which state transitions $s^{t+1}$ reflect cycles of action and counteraction.  

A taxonomy by \emph{actors} determines the set of players $N$. Nation-states, proxies, non-state groups, and insiders all enter with different information, resources, and objectives. State-state interactions are often modeled as \emph{strategic-form or signaling games}, while insider threats require \emph{principal-agent models} to capture hidden actions and incentive misalignments within organizations \cite{zhang2019insurance,zhu2025generative,zhu2025revisiting}.  

In terms of \emph{objectives}, each campaign type changes the payoff functions $u_i$. Attritional warfare emphasizes resource exhaustion, subversive warfare shifts utilities toward reputation and trust, psychological warfare incorporates fear or deterrence, and economic warfare links payoffs to financial disruption. The equilibrium concept $E$ depends on what players value most.  

Finally, classification by \emph{methods} aligns with the appropriate game structures under uncertainty. Sabotage and disruption are modeled with \emph{stochastic or differential games} to capture cascading effects. Espionage fits naturally into a \emph{Bayesian game}, with hidden types and incomplete information. Disinformation and propaganda are framed as \emph{signaling or information design games} \cite{yang2023designing,zhang2021equilibrium,zhang2021informational}, where one player manipulates beliefs to influence the other’s response.  

Thus, a game-theoretic taxonomy does more than describe. It systematically links capabilities, actors, objectives, and methods to specific classes of games. This provides explanatory clarity, why adversaries behave as they do, and predictive power, what equilibria are likely to emerge, making it a practical tool for analysts, policymakers, and cybersecurity professionals.  

\begin{table}[h!]
\centering
\small
\renewcommand{\arraystretch}{1.2}
\begin{tabular}{|p{2.3cm}|p{3.5cm}|p{5.3cm}|}
\hline
\textbf{Category} & \textbf{Examples} & \textbf{Associated Game Models} \\
\hline
Capabilities & Asymmetric, Symmetric, Escalatory & Security games with resource imbalance; Zero-sum matrix games ($u_d=-u_a$); Repeated/stochastic games with state transitions. \\
\hline
Actors & States, Proxies, Non-State Groups, Insiders & Strategic-form or signaling games; Bayesian games (incomplete information); Principal-agent models (hidden actions). \\
\hline
Objectives & Attritional, Subversive, Psychological, Economic & Payoff functions $u_i$ emphasize resource depletion, reputation, deterrence, or financial disruption. \\
\hline
Methods & Sabotage, Espionage, Disinformation, Disruption & Stochastic/differential games (cascading effects); Bayesian games (hidden types); Signaling/information design (belief manipulation). \\
\hline
\end{tabular}
\caption{A game-theoretic taxonomy of cyber warfare: actors define players $N$, capabilities shape actions $A_i$, objectives define payoffs $u_i$, and methods select the appropriate game structure.}
\label{tab:cyber-taxonomy}
\end{table}

\section{Case Study: China and Taiwan Cyber Warfare}

This section develops a hypothetical case study to illustrate how a multi-echelon cyber campaign can unfold and how game-theoretic models help analyze the interaction between China and Taiwan.

\subsection{Current State}

The cyber conflict between China and Taiwan has escalated alongside broader geopolitical tensions and increasingly reflects characteristics of hybrid warfare \cite{solmaz2024chinas}. Episodes around politically sensitive events, such as high-level visits or elections, often coincide with surges in cyber operations. These include DDoS attacks, misinformation campaigns, cyber-espionage, and targeted infrastructure probing, consistent with broader global threat trends documented in recent cybersecurity assessments \cite{enisa2025threat, microsoft2023volt}.

For example, during Nancy Pelosi’s visit to Taiwan in August 2022, cyberattacks hijacked digital billboards to display anti-Pelosi messages, revealing vulnerabilities in Taiwan’s public information infrastructure. Similarly, in April 2023, when Taiwanese President Tsai Ing-wen met U.S. House Speaker Kevin McCarthy, China coupled military maneuvers with coordinated cyber activities, signaling an integrated cross-domain pressure strategy \cite{solmaz2024chinas}.

Other incidents, such as the severing of undersea cables to Taiwan’s Matsu Islands in February 2023, underscored the fragility of communication infrastructure and the strategic importance of redundancy. In response, Taiwan has accelerated resilience initiatives, including emergency microwave relays and the development of a low-Earth-orbit (LEO) satellite communications ecosystem \cite{wang2026taiwan, starlink2023taiwan}. These efforts reflect a broader shift toward diversified and space-enabled connectivity to mitigate geopolitical risk.

Institutionally, Taiwan has strengthened its governance framework through the establishment of the Ministry of Digital Affairs (MODA) and the expansion of agencies such as the National Institute of Cyber Security \cite{moda2022, nics2023}. These reforms align with the objectives outlined in Taiwan’s National Cyber Security Program (2021–2024), which emphasizes infrastructure protection, public–private coordination, and strategic resilience planning \cite{national_cyber_security_program_taiwan_2021_2024}. Collectively, these developments signal a transition from reactive cybersecurity measures toward a comprehensive resilience-oriented national strategy.

\subsection{RedCyber Warfare: Coordinated Cyber Campaign on Taiwan}

As visualized in Fig.~\ref{fig:timeline-bw}, \emph{RedCyber} unfolds as a stage-structured game that pushes from \emph{policy} to \emph{reconnaissance}, \emph{disruption}, \emph{escalation}, \emph{sustained assault}, and \emph{strategic messaging}, with solid arrows showing top-down propagation of objectives/budgets and dashed arrows showing bottom-up feedback from tactical/technical outcomes to higher echelons. At each stage $k$, China’s actions $a_a^k$ and Taiwan’s defenses $a_d^k$ co-evolve via
\[
s^{k+1}\!\sim\!T(s^k,a_d^k,a_a^k;\lambda,\theta), 
\qquad
U^O_d=\sum_{k=1}^K u^O_d(s^k,a_d^k,a_a^k,\lambda),
\]
where $\lambda$ is the \emph{deception index} (which deception mechanism is active: honeypots, decoy credentials, canaries, tarpits), and $\theta$ is the \emph{technology lever} (LLM/agentic AI tooling, automated detection, CPS hardening). Larger $\lambda$ expands belief-shaping options; improved $\theta$ lowers the detection latency and alters feasible action sets, changing the equilibria between phases.

\para{Policy echelon and the feasible downstream game.}
Months before a visible crisis, Beijing’s doctrine blends intrusion, economic pressure, and information ops; Taipei counters by setting weights and budgets $(w,B)$, committing to deception portfolios $\lambda$, and activating alliances that provide surge capacity. In Fig.~\ref{fig:timeline-bw}, this is the \emph{Policy} node ($k{=}0$), which pushes a feasible game downstream by constraining the resources and timing available to the \emph{Strategic} and \emph{Operational} nodes. Formally, the policy equilibrium $E_{\mathrm{policy}}$ parameterizes $\mathcal{G}_{\mathrm{strategic}}$ and the budget sets therein, consistent with cross-echelon meta-game design \cite{yang2025multires,huang2020dynamic,chen2019dynamic}. Empirically, Taiwan’s establishment of MODA and associated resilience programs provides an institutional anchor for $(w,B,\lambda)$ and $\theta$ updates \cite{moda-site}. 

\para{Reconnaissance as signaling with evidence.}
APT teams exploit zero-days in edge devices and conduct tailored phishing of municipal IT staff; sock puppets seed test narratives on social platforms to probe which themes resonate. Defenders plant decoy credentials and canary tokens to induce observable errors and generate \emph{micro-evidence} for attribution without revealing coverage. This corresponds to \emph{asymmetric-information/signaling-with-evidence} games (the \emph{Recon} node in Fig.~\ref{fig:timeline-bw}), where the attacker learns hidden types and the defender manipulates observability \cite{pawlick2018leaky,zhang2018hypothesis,huang2019adaptive,huang2020farsighted,huang2021duplicity}. Episodes around sensitive politics illustrate the timing: during Speaker Pelosi’s August 2022 visit, Taiwan experienced surges in DDoS and screen hijacks of public displays \cite{reuters-pelosi-2022,abc-pelosi-2022,cyberscoop-2022}. 

\para{Disruption as a sequential CPS game.}
With footholds established, malware on industrial control systems (ICS) triggers localized outages while DDoS floods emergency portals; fabricated outage maps amplify panic. Taiwan’s counter is to island microgrids, shift traffic to prepositioned static mirrors and CDN, and create emergency communications (ERCV / LEO terminals) to maintain command and control. These interactions match \emph{sequential attack-defense} games on the CPS (the \emph{Disruption} node), with hybrid/stochastic models that capture cascades and rapid response \cite{miao2018hybrid,zhao2020semiNP,chen2019dynamic}. The link to practice is visible in the response to the Hualien earthquake of Taiwan on April 3, 2026, where MODA deployed the first Mobile Emergency Network Vehicle and later airlifted LEO satellite terminals to restore connectivity, a concrete increase in $\theta$ that steepens attacker cost curves in $T(\cdot)$ \cite{moda-ercv-2024a,moda-ercv-2024b}.

\para{Escalation as Stackelberg with information design.}
As elections approach, credential-stuffing and spoofed ``mirror'' results appear; deepfakes and forged documents attempt to delegitimize candidates; semiconductor and transport disruptions compound shocks. Taiwan activates real-time audits, paper backups, golden-image rollbacks, and transparent livestreams. The \emph{Escalation} node is well captured as a \emph{Stackelberg} game: the attacker leads with politically/economically coupled moves; the defender responds under resource constraints \cite{liu2025stackrisk}. Information-design tools formalize the overt/covert counter-messaging trade-offs \cite{li2023price}. In practice, cyber and kinetic signaling have been tightly coupled around high-visibility diplomacy and exercises \cite{icg-2023}. Cloud measurement shows Taiwan-bound DDoS spiking around election periods, aligning with timing games that exploit attention cycles \cite{cloudflare-q4-2023,cloudflare-q1-2023}.

\para{Sustained assault as a repeated game with incentives.}
The tempo settles into weekly breach dumps, payment hiccups, and refresh spearphishing to re-seed access. Taiwan rotates keys (including PQC pilots), segments high-value data, automates fraud detection, and uses insurance/contracting to share risk. This is the \emph{Sustained Assault} node, can be modeled as a \emph{repeated game} in which resilience investments and incentive instruments depress the attacker’s marginal returns \cite{zhu2020crosslayer,zhang2019insurance,liu2025stackrisk}. Physical infrastructure remains a live risk channel; repeated damage to undersea cables near the Matsu islands has forced backup communication and hardened procedures, illustrating how exogenous shocks perturb $T(\cdot)$ and the feasible action sets \cite{reuters-matsu-2025,ap-matsu-2023}.

\para{Strategic messaging as signaling/information design.}
Finally, partial restoration of services is overlaid with inevitability narratives; embassy spoofs and ``relief phishing'' siphon credentials. Taiwan responds with cryptographically signed fact-checks, international amplification of verified messages, and rapid take-down of coordinated inauthentic behavior. The \emph{Strategic Messaging} node is a \emph{signaling (information-design)} game in which credibility, transparency, and commitment shape equilibrium beliefs and blunt coercion \cite{li2023price,liu2023dim}. 

\para{Cross-echelon coupling.}
Throughout, the deception index $\lambda$ and technology lever $\theta$ (annotated above the timeline in Fig.~\ref{fig:timeline-bw}) move the frontier of feasible actions and detection latencies. In equilibrium terms, $E_{\text{policy}}$ sets budgets, alliances, and doctrine that parameterize $\mathcal{G}_{\text{strategic}}$; $E_{\text{strategic}}$ allocates resources/timing that constrain $\mathcal{G}_{\text{operational}}$; operational outcomes filter feasible tactics and belief states; and tactical/technical outcomes feed \emph{back} (dashed arrows) to shift higher-level equilibria by changing costs and beliefs. Advances in $\theta$, notably LLM/agentic-AI defenses for detection, reasoning, and counter-deception, have been modeled directly as LLM-Nash/Stackelberg games and deception timing games; these mechanisms compress attacker windows and reshape upstream plans \cite{zhu2025llmnash,zhu2025llmstack,li2024symbiotic,yang2025deceive}. In short, RedCyber behaves as \emph{meta-games}: The equilibrium of each node depends on and perturbs the rest, with real-world incidents providing clear calibration points for $T(\cdot)$, costs and beliefs.

\begin{figure}[t]
\centering
\begin{tikzpicture}[
  >=Latex,
  line cap=round,
  node distance=12mm and 10mm,
  card/.style={draw=black, rounded corners=1.5pt, minimum width=3.9cm, minimum height=1.6cm, align=left, fill=white},
  date/.style={font=\bfseries\small},
  small/.style={font=\scriptsize},
  spine/.style={very thick, draw=black},
  stem/.style={thick, draw=black},
]

\colorlet{accentc}{black}



\draw[spine] (-1.0,0) -- (10.5,0);

\coordinate (A) at (0.6,0);
\coordinate (B) at (2.6,0);
\coordinate (C) at (4.6,0);
\coordinate (D) at (6.6,0);
\coordinate (E) at (8.6,0);

\newcommand{\eventup}[5]{%
  \draw[stem] (#2) -- ++(0,6mm) coordinate (p-#1);
  \node[card, anchor=south] (b-#1) at (p-#1) {};
  \node[date, text=black, anchor=north west]  at ($(b-#1.north west)+(2mm,-2mm)$) {#3};
  \node[small, anchor=north west]             at ($(b-#1.north west)+(2mm,-7mm)$) {\textbf{#4}};
  \node[small, anchor=north west]             at ($(b-#1.north west)+(2mm,-11mm)$) {#5};
}
\newcommand{\eventdown}[5]{%
  \draw[stem] (#2) -- ++(0,-6mm) coordinate (p-#1);
  \node[card, anchor=north] (b-#1) at (p-#1) {};
  \node[date, text=black, anchor=south west] at ($(b-#1.south west)+(2mm,2mm)$) {#3};
  \node[small, anchor=south west]            at ($(b-#1.south west)+(2mm,7mm)$) {\textbf{#4}};
  \node[small, anchor=south west]            at ($(b-#1.south west)+(2mm,11mm)$) {#5};
}

\eventdown{pol}{A}{Jan--Mar 2026}{Policy \& Preparation ($k{=}0$)}{Goals, budgets, alliances}
\eventup  {rec}{B}{Apr--Sep 2026}{Reconnaissance ($k_1$)}{Asymmetric info; $\lambda$ decoys/canaries}
\eventdown{dis}{C}{Oct--Dec 2026}{Initial Disruption ($k_2$)}{CPS/DDoS; $\theta$ shifts $T(\cdot)$}
\eventup  {esc}{D}{Jan--Mar 2027}{Escalation ($k_3$)}{Leader--follower responses}
\eventdown{sus}{E}{Apr 2027--ongoing}{Sustained Operations}{Repeated play; incentives/insurance}

\end{tikzpicture}
\caption{Stage-structured timeline of the \emph{RedCyber} campaign against Taiwan. 
Each node illustrates a phase of the campaign, policy preparation, reconnaissance, disruption, escalation, and sustained operations, with the deception index $\lambda$   and technology lever $\theta$ annotated as cross-cutting parameters that shift detection latencies and feasible action sets in $T(\cdot)$.}

\label{fig:timeline-bw}
\end{figure}

\section{Conclusions}
The advent of cyber warfare has introduced unprecedented challenges that require innovative frameworks to understand and manage conflicts in the digital domain. Game theory emerges as a critical tool that offers structured methodologies to model, analyze and predict interactions on different scales of cyber operations. By integrating concepts such as equilibrium analysis, dynamic adaptation, and resource optimization, game-theoretic approaches provide actionable insights into adversarial dynamics and strategic planning. These methods enable policymakers, strategists, and defenders to navigate the complexities of cyber warfare effectively, aligning localized actions with broader objectives.

Furthermore, game theory's ability to address uncertainties, manage risks, and foster cooperation highlights its relevance in both offensive and defensive cyber operations. Whether through dynamic games that model escalation cycles or multi-scale frameworks that link tactical decisions to strategic outcomes, game theory empowers actors to adapt and innovate in a rapidly evolving landscape. As cyber warfare continues to evolve, the integration of game-theoretic principles will be indispensable to maintain resilience, achieve operational superiority, and ensure global security in the face of emerging threats.

\vspace{10mm}

\bibliographystyle{abbrv}
\bibliography{reference}

\end{document}